\documentclass[twocolumn, prb, superscriptaddress]{revtex4-2}

\usepackage[utf8]{inputenc}
\usepackage{graphicx}
\usepackage{amsmath}
\usepackage{hyperref}
\usepackage{amssymb}
\usepackage{booktabs} 
\usepackage{siunitx} 
\usepackage{float}
\usepackage{xcolor}
\usepackage{subcaption} 
\usepackage{diagbox}
\usepackage{overpic}

\newcommand{\beq}{\begin{eqnarray} }
\newcommand{\eeq}{\end{eqnarray} }
\newcommand{\Beq}{\begin{eqnarray*} }
\newcommand{\Eeq}{\end{eqnarray*} }
\newcommand{\Bmat}{\left(\begin{matrix}}
\newcommand{\Emat}{\end{matrix}\right)}
\newcommand{\up}{\uparrow}
\newcommand{\dn}{\downarrow}

\usepackage{etoolbox}
\AtBeginEnvironment{thebibliography}{%
	\small
	\def\url#1{}%
	\def\doi#1{}%
}

\setcitestyle{super,sort&compress}

\begin{document}

\title{Deconfined Gapless Phases and criticalities in Shastry-Sutherland Antiferromagnet} 

\author{Lvcheng Chen}
\affiliation{Department of Physics, Fudan University, Shanghai 200433, China}
\affiliation{School of Physics and Beijing Key Laboratory of Opto-electronic Functional Materials and Micro-nano Devices, Renmin University of China, Beijing, 100872, China}
\affiliation{Key Laboratory of Quantum State Construction and Manipulation (Ministry of Education), Renmin University of China, Beijing, 100872, China}
\affiliation{Key Laboratory for Quantum Materials of Zhejiang Province, School of Science, Westlake University, Hangzhou 310024, China}

\author{Zheng-Xin Liu}
\email{liuzxphys@ruc.edu.cn}
\affiliation{School of Physics and Beijing Key Laboratory of Opto-electronic Functional Materials and Micro-nano Devices, Renmin University of China, Beijing, 100872, China}
\affiliation{Key Laboratory of Quantum State Construction and Manipulation (Ministry of Education), Renmin University of China, Beijing, 100872, China}

\date{\today}

\begin{abstract}
Antiferromagnets on the Shastry-Sutherland lattice have attracted lots of research interest due to the possible existence of deconfined criticality. In the present work, we study the $J_1$-$J_2$-$J_r$ model using Variational Monte Carlo (VMC) method, where $J_1$, $J_2$, $J_r$ stand for the nearest-neighbor, next nearest neighbor and ring exchange interactions respectively. An empty plaquette (EP) phase with spontaneous mirror symmetry breaking is reproduced. However,  the EP phase in the VMC approach is $Z_2$ deconfined and have Majorana-type gapless spinon excitations, which is qualitatively different from the EP phase in literature. The central observation of the present study is the gapless $Z_2$ Quantum spin liquid phase 
resulting from the competition between the EP phase, the full plaquette (FP) phase and the  antiferromagnetic N\'eel phase. While the phase transition from the $Z_2$ QSL phase to the EP phase is likely of Landau-Ginzburg type, the continuous transitions to the confined FP and N\'eel phases are exotic and need to be further explored.

\end{abstract}

\maketitle

 \section{Introduction} The antiferromagnic (AFM) $J_1$-$J_2$ Heisenberg model 
 on the Shastry-Sutherland (SS) lattice\cite{Shastry1981Exact} is strongly frustrated and hosts several interesting phases, including the dimer-singlet phase, the empty plaquette (EP) singlet phase, the N\'eel ordered phase and a possible quantum spin liquid phase (QSL). The dimer phase is a symmetric phase with short-range entanglement, the plaquette phase breaks the mirror reflection symmetry and the N\'eel phase breaks the spin rotation symmetry. While the Landau-Ginzburg theory prevents a direct continuous phase transition between two phases breaking different symmetries, it was proposed that the transition between the plaquette phase and the N\'eel ordered phase belongs to the exotic deconfined critical point (DQCP) \cite{Senthil_DQCP_science, Senthil_DQCP_prb, wangchong_duality}.  Unlike the QSL which is a stable deconfined phase with finite volume in the phase space, the DQCP is an exotic point or line 
 \cite{Sandvik_2007, Song_2025, wangcong_PANS}. Thereafter, the antiferromagnets on the SS lattice had attracted lots of research interest\cite{YouPRX,Sandvik_SS,series_expansion_SS,iPEPS_SS,XieZhiyuan,Wangfa,Wangling,WanglingED,KaiPhillipSchmidt}. 

On the experimental side, the material SrCu$_2$(BO$_3)_2$ is an ideal candidate antiferromagnet with deformed SS lattice geometry, in which a pressure and magnetic-field induced phase transition proximate to a DQCP was observed with nuclear magnetic resonance technique\cite{NP2017,WeiqiangYu}. SS antiferromagnets with strong spin-orbital couplings have also been reported recently\cite{PhysRevB.110.144445, liu2024theoryrareearthkramersmagnets, li2024spinonsnewshastrysutherlandlattice, Pula_2024}.
Theoretically, the phase diagram of the SS antiferromagnetic Heisenberg model and the nature of the phase transitions are still under debate. For instance,  Schwinger boson mean-field theory (SBMFT)\cite{Wangfa} suggest that the intermediate phase between the Dimer phase and the N\'eel phase is a symmetric gapped $Z_2$ spin-liquid  (for $0.66<J_1/J_2<0.71$, here $J_1, J_2$ stands for the AFM Heisenberg exchange interactions on the nearest and next nearest neighbor bonds) instead of a plaquette-singlet phase. But more accurate numerical computations indeed confirm the existence of the EP phase (valence bond solid order in the empty plaquettes having no diagonal bonds) in the intermediate region. The transition between the EP phase and the N\'eel phase was proposed to be a DQCP with emergent O(4) symmetry \cite{YouPRX}. This proposal was supported by tensor network studies\cite{Guzhengcheng_nospinliquid}, which suggests that there is no spin liquid in the intermediate region. 
It was also proposed  that there may exist a deconfined phase (instead of a critical point), {\it i.e.} a QSL, between the EP phase and the N\'eel phase\cite{Wangling, iPEPS_SS, WanglingED, FRG, iPEPS_2}. Furthermore, the spin wave bands in the N\'eel phase was identified with an altermagnetism\cite{Yurong_alter,VMC_alter}, and the emergent O(4) symmetry was 
supported by the elementary excitations close to the transition point. Some other numerical work suggested that the phase transition may be of  weak first order\cite{iPEPS_SS,YouPRX,GuoWenAn_JQ}, and a ring-exchange interaction can probably turn the first-order transition into a continuous one\cite{XieZhiyuan}.

\begin{figure}
	\includegraphics[width=.98\linewidth]{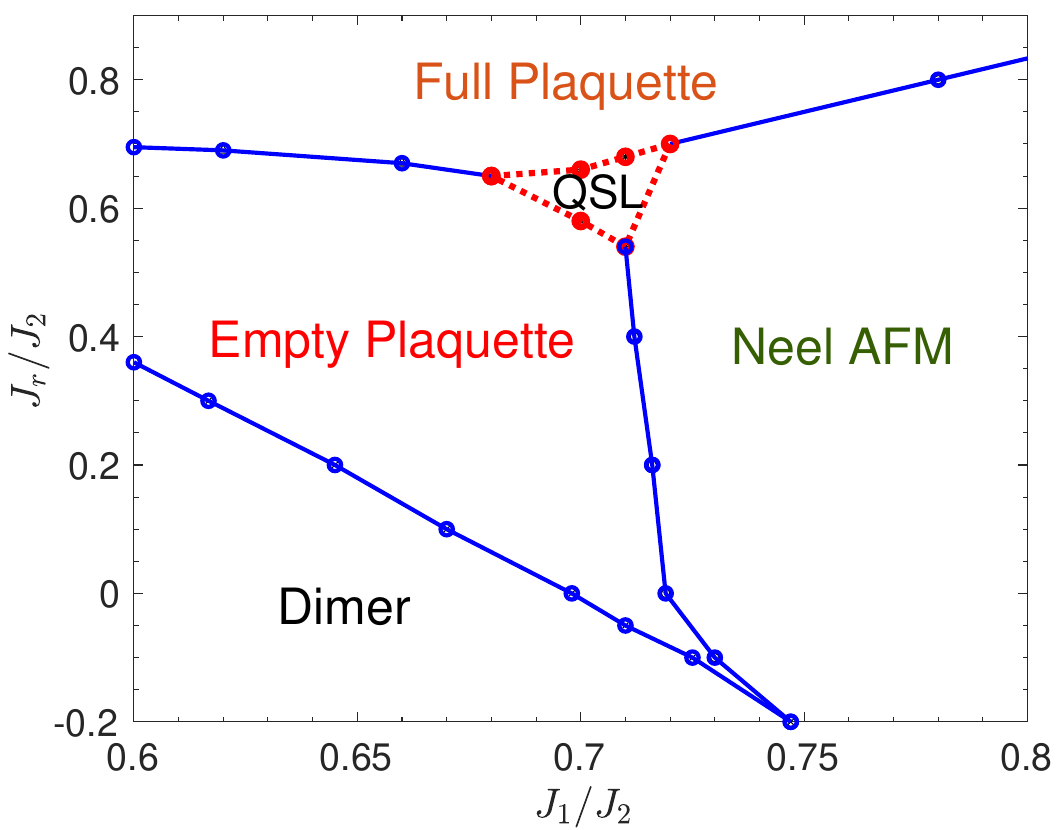}
	\caption{ Phase diagram of \( J_1 \)-\( J_2 \)-\( J_r \) model on the Shastry-Sutherland lattice. The solid/dotted lines represent first-order/continuous phase transitions. }\label{fig:phasedgrm}
\end{figure}

To unveil the nature of the interesting intermediate phases, we study a AFM model on the SS lattice using variational Monte Carlo (VMC) method. By including the Heisenberg and ring exchange interactions, we obtain a rich phase diagram (see Fig.\ref{fig:phasedgrm}) composed of a dimer phase, two plaquette phases, a N\'eel phase and a $Z_2$ QSL phase. A counterintuitive observation is that the EP phase is a $Z_2$ deconfined gapless phase having four Majorana cones in the excitation spectrum. The only difference between the EP phase and the  $Z_2$ QSL phase is that the former spontaneously breaks the mirror symmetry.
Furthermore, the phase transitions between the QSL and the neighboring symmetry breaking phases are found to be continuous. We propose critical scenarios for these interesting  transitions. 

The rest part of the paper is organized as follows. The microscopic model and the VMC are introduced in section \ref{sec:MM}, while the phase diagram and the nature of each phase are presented in section \ref{sec:PD}, and the scenarios of the continuous phase transitions are discussed in section \ref{sec:PT}. Section \ref{sec:conc} is devoted to the conclusion and discussion. 

\section{ The model and VMC method} \label{sec:MM} 
\begin{figure}[t]
	\centering
	\	\includegraphics[width=1\linewidth]{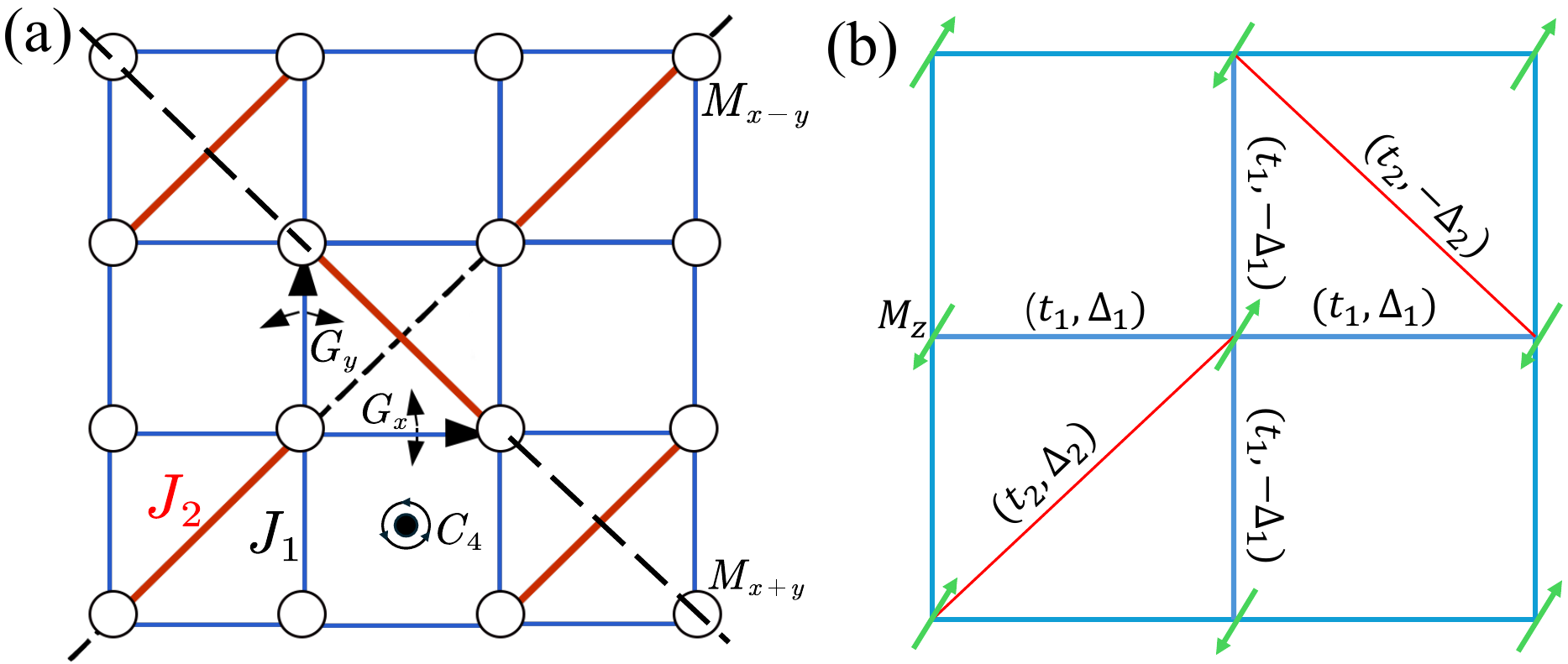}
	\caption{(a) The SS lattice model with the translation $T_x, T_y$, rotation $C_4$, mirror $M_{x\pm y}$, and glide reflection $G_x, G_y$ symmetries. (b) The spin liquid ansatz Z2Azz13SS with $(t_{ij},\Delta_{ij})$ standing for the hopping and pairing parameters on the corresponding bonds. }\label{fig:SSmodel}
\end{figure}

We consider the following $J_1$-$J_2$-$J_r$ model on the SS lattice, 
\beq\label{J1J2J4}
	H=J_1 \sum_{\langle i j\rangle} \!\mathbf{S}_i \cdot \mathbf{S}_j \!+\! J_2\!\! \sum_{\langle\langle i j\rangle\rangle}\! \mathbf{S}_i \cdot \mathbf{S}_j \!-\! J_r \!\!\!\!\sum_{i j k l \in {\rm [FP]}} \!\!\! Q_r(ijkl), 
\eeq
where $\langle ij \rangle$ labels nearest neighbors, $\langle\langle ij \rangle\rangle$ stands for next-nearest-neighbor bond along the diagonal directions (see Fig.~\ref{fig:SSmodel}(a)), and $Q_r$ is the 4-spin ring-exchange interaction on the full-plaquettes (FP) containing the diagonal bonds\cite{XieZhiyuan}, namely,  
\beq
Q_r(ijkl) = &&\Big[  \left( \mathbf{S}_i \cdot \mathbf{S}_j \right) \left( \mathbf{S}_k \cdot \mathbf{S}_\ell \right) + \left( \mathbf{S}_j \cdot \mathbf{S}_k \right) \left( \mathbf{S}_\ell \cdot \mathbf{S}_i \right) \notag\\ 
		&& - \left( \mathbf{S}_i \cdot \mathbf{S}_k \right) \left( \mathbf{S}_j \cdot \mathbf{S}_\ell \right) \Big].
\eeq
The SS lattice has a non-symmorphic space group symmetry containing the following symmetries:
translation operations $T_x, T_y$,  4-fold rotation $C_4$, glide reflection, $G_x=\{M_x | t_x\}, G_y=\{M_y | t_y\}$, and mirror reflections $M_{x\pm y}$, where $t_x=T_x^{1/2}, t_y=T_y^{1/2}$.

In the following, we study the above model using VMC method. The VMC approach is based on the fermionic parton representation, where the spin operators are presented as $\mathbf{S}_i=\frac{1}{2} \sum_{\alpha,\beta=\up\dn} f_{i \alpha}^{\dagger} \pmb{\sigma}_{\alpha \beta} f_{i \beta}$ under the single occupancy constraint $\sum_\alpha f_{i \alpha}^{\dagger} f_{i \alpha}=1$. A general mean-field Hamiltonian for the SS model then reads
\beq
	H_{\rm MF} &\!=\!& \sum_{ij} \Big(t_{ij } \sum_\alpha f_{i\alpha}^\dagger f_{j\alpha} 
	+ \Delta_{ij} f_{i\uparrow}^\dagger f_{j\downarrow}^\dagger + \text{h.c.} \Big)\notag \\
	&& + \!\sum_i\!\Big[\lambda_z \!\sum_\alpha f_{i\alpha}^\dagger f_{i\alpha} 
	+ M_z \big( f_{i\uparrow}^\dagger f_{i\uparrow} - f_{i\downarrow}^\dagger f_{i\downarrow} \big)\Big], \label{eq:HMF}
\eeq
with $t_{ij}$ the hopping parameter ($t_1$ for nearest neighboring bonds and $t_2$ for diagonal bonds), $\Delta_{ij}$ the pairing parameter ($\Delta_1$ for nearest neighboring bonds and $\Delta_2$ for diagonal bonds, see Fig.\ref{fig:SSmodel}(b)), $\lambda_z$ the Lagrangian multiplier, and $M_z$ the background field due to spontaneous magnetization. For the states with plaquette orders, we will introduce more parameters later to characterize the bond modulation of the entanglement intensity.

The spin operators can also be expressed as $S^\mu_i = \frac{1}{4} \text{Tr}(\Psi_i^\dagger \sigma^\mu \Psi_i)$ with 
$\Psi_i = 
\begin{pmatrix}
f_{i\uparrow} & f_{i\downarrow}^\dagger \\
f_{i\downarrow} & -f_{i\uparrow}^\dagger
\end{pmatrix}
$ \cite{anderson_su2, LZN_PRB10}. 
The spin operators are invariant under local transformations, 
$\Psi_i \to \Psi_i G_i, \quad G_i \in SU(2),$
reflecting a $SU(2)$ gauge structure of the fermion representation.
The mean-field Hamiltonian can also be written as:
\begin{equation}\label{eq:Hmf} 
H_\text{MF} = {1\over4}\sum_{ij} \text{Tr}(\Psi_i u_{ij} \Psi_j^\dagger),
\end{equation}
with $u_{ij} = i {\rm Im}(\chi_{ij}) \tau^0 + {\rm Re}(\chi_{ij})\tau^z + {\rm Re}(\Delta_{ij})\tau^x + {\rm Im}(\Delta_{ij})\tau^y$ and $u_{ii} = \lambda_i^z \tau^z + \lambda_i^x \tau^x + \lambda_i^y \tau^y$, where $\tau^0$ is the identity matrix and $\tau^{x,y,z}$ the Pauli matrices generating the $SU(2)$ gauge group. We fix the gauge such that $\lambda_x=\lambda_y=0$ and determine the value of $\lambda_z$ by variation.
The invariant gauge group (IGG) of the mean field Hamiltonian (\ref{eq:Hmf}) is either $U(1)$ or $Z_2$, depending on the values of the pairing parameters. Furthermore, the symmetry group  is generally the extension of the space-time symmetry group by the IGG, which is called the projective symmetry group (PSG)\cite{wen_PSG}. 

The Gutzwiller projected ground state of the above Hamiltonian, namely 
$$
|\Psi(\pmb R)\rangle = P_G|{\rm GS}(\pmb R)\rangle_{\rm mf},
$$
which provides trial wave function for the spin model (\ref{J1J2J4}), where $|{\rm GS}(\pmb R)\rangle_{\rm mf}$ is the ground state of the trial Hamiltonian (\ref{eq:HMF}), $\pmb R=(\chi_1, \chi_2, \Delta_1, \Delta_2, \lambda_{x,y,z}, M_z, ...)$ are variational parameters and $P_G$ stands for the Gutzwiller projection that enforces the single occupancy constraint.
The energy of the trial state 
\Beq
E_{\rm trial} = {\langle\Psi(\pmb R) |H|\Psi(\pmb R)\rangle\over\langle\Psi(\pmb R)|\Psi(\pmb R)\rangle} = \sum_{\alpha} \rho(\alpha) \Big[\sum_\beta {f(\beta) \over f(\alpha)} H_{\alpha\beta}\Big]
\Eeq
can be computed using Monte Carlo sampling, where $f(\alpha) = \langle \alpha |\Psi(\pmb{R}) \rangle$ is the amplitude of the Gutzwiller projected wave function and $\rho(\alpha) =\frac{|f(\alpha)|^2}{\sum_\beta |f(\beta)|^2}$ is the normalized probability. By minimizing the energy $E_{\rm trial}(\pmb R)$, we can obtain the optimal parameters $\pmb R$ for every given $J_1/J_2, J_r/J_2$, and then figure out the phase diagram. 

Throughout all the interaction regions, nonzero values of pairing terms $\Delta_1, \Delta_2$ are energetically favored in our VMC calculation, indicating that the IGG is $Z_2$. Furthermore, the confinement/deconfinement of $Z_2$ gauge field can also be detected by the ground state degeneracy (GSD) on a torus. The information of GSD can be inferred by calculating the overlap of wave functions in different topological sectors \cite{testingZ2TO, WangLiu_KiteavSL, Liu_U1SPT}. The four topological sectors are distinguished by the $Z_2$ gauge flux in the two holes of the torus, which are reflected by the periodic or anti-periodic boundary conditions along the $x$- and $y$-directions for the fermions in the mean field Hamiltonian (\ref{eq:HMF}), namely, $|\psi_1\rangle=|++\rangle,  |\psi_2\rangle=|+-\rangle, |\psi_3\rangle=|-+\rangle, |\psi_4\rangle=|--\rangle$ where $+/-$ stands for periodic/anti-periodic boundary condition along one of the two directions. Using Monte Carlo method, one can calculate the fidelity matrix $F$ formed by the overlap of the above 4 wave functions, 
$$
F_{ab}=\left. \langle \psi _a | \psi _b \right. \rangle
=\frac{1}{C}\sum_{\alpha}{\rho_a}(\alpha ) \frac{f_b(\alpha )}{f_a(\alpha )},
$$
with $a,b=1,2,3,4$, $\rho_a(\alpha) =\frac{|f_a(\alpha)|^2}{\sum_\beta |f_a(\beta)|^2}$ and  $C = \sqrt{\frac{\sum_\beta{|f_{b}(\beta)|^{2}}}{\sum_\alpha{|f_{a}(\alpha)|^{2}}}}$ a normalization constant. In the $Z_2$ deconfined phase, the GSD is 4, namely, the above 4 states are orthogonal to each other, hence the eigenvalues of the fidelity matrix $F$ should be (1,1,1,1); in the $Z_2$ confined phase, GSD=1, so the 4 states are the same, hence the eigenvalues of $F$ should be (0,0,0,4).  Detailed discussion can be found in App.\ref{app:overlap}.

\section{Phase diagram} \label{sec:PD}


By performing VMC calculations of the $J_1$-$J_2$-$J_r$ model with $12\times12$ sites, we obtain the phase diagram with five distinct phases, as shown in Fig.\ref{fig:phasedgrm}. Three of the the phases appear at the $J_1$-$J_2$ model with $J_r=0$, namely the dimer phase at $J_1/J_2<0.698$,  the EP phase in the interval $0.698<J_1/J_2<0.719$, and the N\'eel phase for $J_1/J_2>0.719$ (see Fig.\ref{fig:J1J2phase diagram} in App.\ref{app:J1J2}). When $J_r$ is negative, the size of the EP phase is reduced and a direct transition from the dimer phase to the N\'eel phase is obtained. On the positive $J_r$ side, a FP phase appears when $J_r$ is large, and a gapless $Z_2$ QSL phase is observed in a small region at the junction of the FP, EP and N\'eel phases. 

\subsubsection{The dimer phase}

The dimer phase is a featureless gapped trivial phase preserving all the lattice symmetries. The ground state is a product of dimer singlets on the diagonal bonds and there is no inter-dimer entanglement. 

In the VMC approach, the dimer phase is characterized by the dominating diagonal hopping or pairing terms and vanishing nondiagonal hopping and pairing parameters. Since the ansatz of the dimer phase is qualitatively distinct from the other phases and the ground state almost remains unchanged in the whole phase, the transitions from the dimer phase to the other phases are of first order. 
\begin{figure}[t]
	\centering
	\	\includegraphics[width=0.85\linewidth]{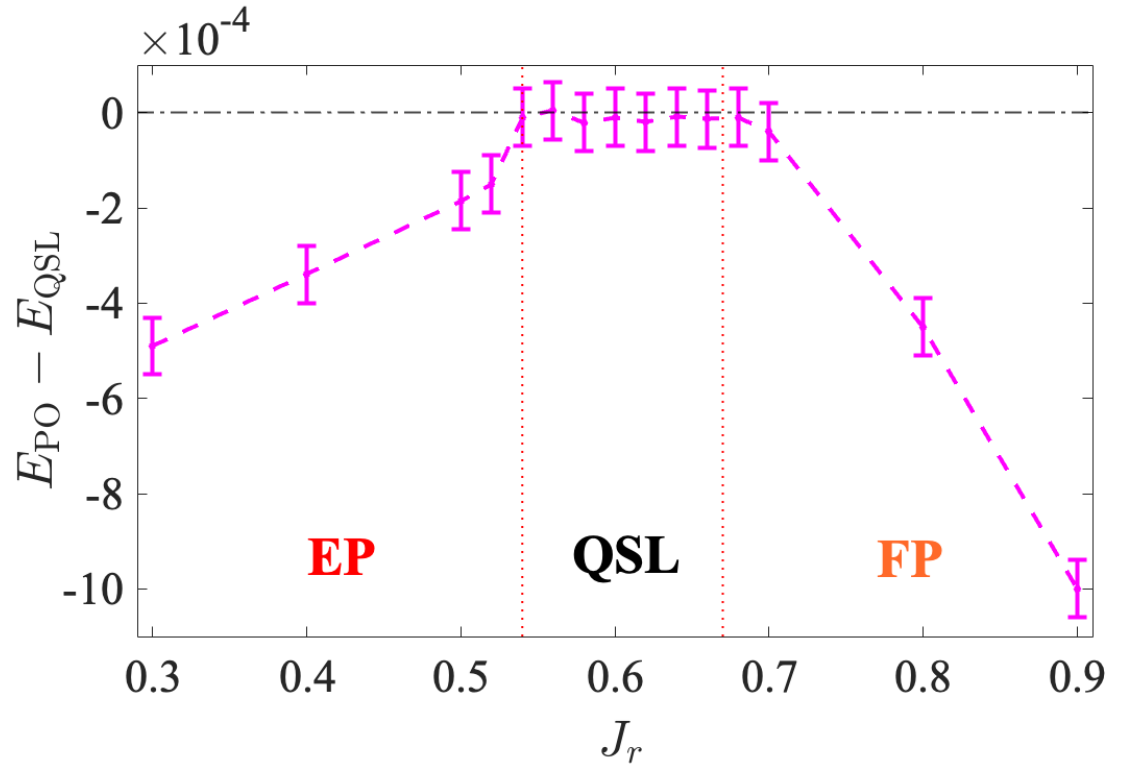}
	\caption{Energy difference between plaquette-ordered states and the QSL Z2Azz13SS, namely $E_{\rm PO}-E_{\rm QSL}$, for $J_1/J_2=0.71$.}
	\label{fig:QSL_energy}
\end{figure}

\subsubsection{The QSL phase}

A QSL phase preserving all of lattice symmetries shows up due to the competition between various classical orders. Symmetry breaking states are very close in energy compared with the QSL according to our VMC calculations. We compare the energies for variational states with or without including symmetry breaking  variational parameters. If the energy of the non-ordered one is equal to the ordered ones within the numerical precision, then we treat the QSL state as the ground state  (see Fig.\ref{fig:QSL_energy} for illustration, and App.\ref{app:mean field rules} for details). 

In the following, we will illustrate that the QSL phase has three features: \\
\indent (I) $Z_2$ gauge symmetry; \\
\indent (II) gapless spinon spectrum;\\
\indent (III) deconfiened $Z_2$ gauge fluctuations.

Firstly, we test various QSL ansatzs in our VMC study. It turns out that the pairing terms (see Eq.(\ref{tilde_uij}) in App.\ref{app:Z2Azz13}), which renders the IGG to be $Z_2$, generally help to lower the energy.  As shown in Tab.~\ref{tab:energy_SL}, the energies of the $Z_2$ SL states are lower than the U(1) SL.

\begin{figure}[t]
	\centering
	\	\includegraphics[width=0.75\linewidth]{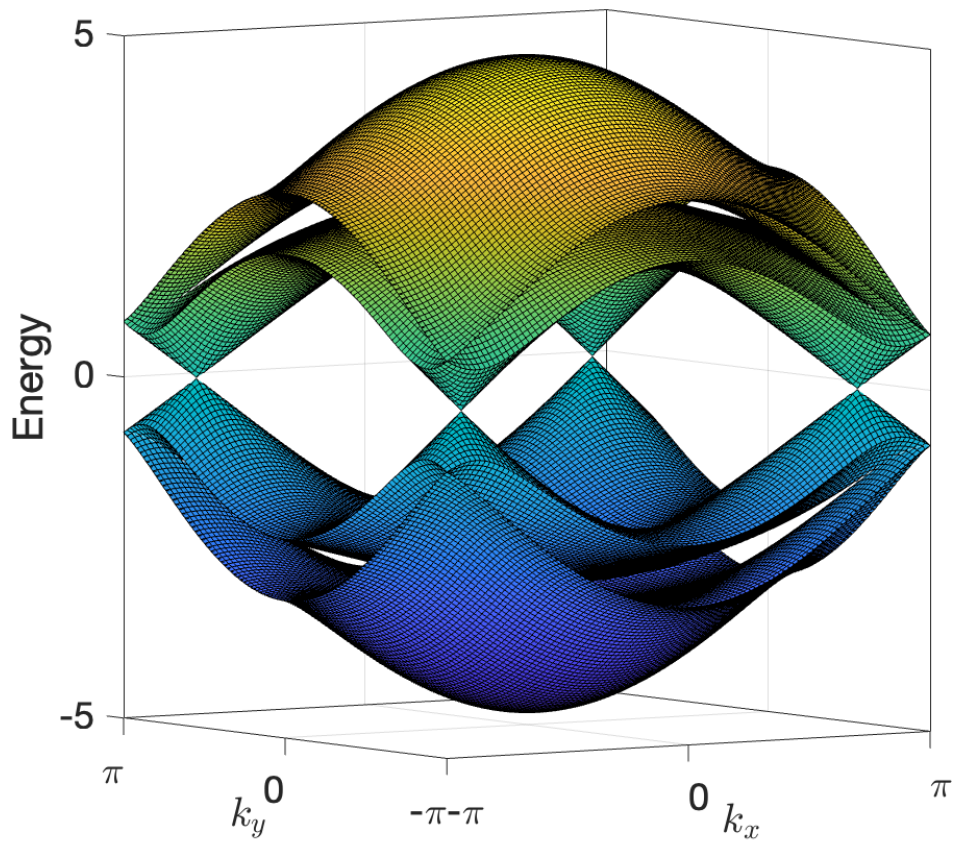}
	\caption{Spinon dispersion of the gapless $Z_2$ QSL phase. Four Majorana cones protected by the combined $IT$ symmetry are locating on the diagonal $(k_x\pm k_y)$-lines.}
	\label{fig:spinon}
\end{figure}

\begin{table}[htbp]	\caption{Comparison of the energies of different QSL ansätze in SL phase with $J_1/J_2=0.7,J_r=0.62$.} 
	\label{tab:energy_SL}
	\centering
	\renewcommand{\arraystretch}{1.2}
	\begin{tabular}{|c|c|c|c|}
		\hline
		ansatz& Z2Azz13SS & $s$-wave SL & U(1) SL \\
		\hline
		Energy per site & -0.47898 & -0.47889 & -0.47820 \\
		\hline
	\end{tabular}
\end{table}

Furthermore, the ansatz with the lowest energy is a gapless $Z_2$ QSL noted as Z2Azz13SS [see Fig.\ref{fig:SSmodel}(b)] in later discussion,
\beq
&&\!\! {u}_{i, i+\hat{x}}=
\Bmat t_1 & \Delta_1 \\
\Delta_1 & -t_1
\Emat,\ \  {u}_{i, i+\hat{y}}=\Bmat
			t_1 & -\Delta_1\\
			-\Delta_1 & -t_1
\Emat \notag \\
&&\!\! {u}_{i, i+\hat{x}+\hat{y}}=\Bmat
			t_2 & \Delta_2 \\
			\Delta_2 & -t_2
\Emat,\ \ {u}_{i, i-\hat{x}+\hat{y}}=\Bmat
			t_2 & -\Delta_2 \\
			-\Delta_2 & -t_2
\Emat. \label{u}
\eeq 
The symmetry group of the ansatz Z2Azz13SS is a  PSG composed of symmetry operations taking the form  $\tilde g = W_g g$, where $g$ is a  space group operation of the SS lattice and $W_g$ is the corresponding gauge transformation. For instance, when fixing the gauge such that $t_2=0$, then the $W_g$ in the PSG are given by
\Beq
&&W_{T_x}(i) = \tau^{0}, \ \  W_{T_y}(i) = \tau^{0}, \\
&&W_{G_x}(i) = (-1)^{i_x+i_y} \mathrm{i}\tau^{y} ,\ \ W_{G_y}(i)= (-1)^{i_x+i_y+1}\mathrm{i}\tau^{y} , \\
&&W_{M_{x+y}}(i) =(-1)^{i_x+i_y}\mathrm{i}\tau^{x}, \  W_{C_4}(i)= \mathrm{i} \tau^z.
\Eeq
More details for the PSG of the QSL ansatz (\ref{u}) are discussed in the App.~\ref{app:PSG}. 

Actually, the above gapless $Z_2$ QSL (\ref{u}) can be deformed from the Z2Azz13 ansatz for the $J_1$-$J_2$ Heisenberg model on square lattice \cite{Sachdev_SL, wen_PSG, HuwenjunSL}. 
Removing part of the diagonal bonds from Z2Azz13 \cite{wen_PSG}, Performing a sequence of gauge transformations (see Fig.\ref{Z2} in App.\ref{app:Z2Azz13}),  one obtains the QSL ansatz Eq.(\ref{u}) which 
is referenced as Z2Azz13SS in the present work. Similar to Z2Azz13, the spinon dispersion in Z2Azz13SS is still gapless, with four Majorana cones locating at $\left( \pi,\pi \right) + \left( \pm \tfrac{1}{2} \tfrac{\sqrt{\Delta_2^2+t_2^2}}{\Delta_1}, \pm \tfrac{1}{2} \tfrac{\sqrt{\Delta_2^2+t_2^2}}{\Delta_1} \right)$ on the $k_x\pm k_y$ diagonal lines shown in Fig.\ref{fig:spinon}. These Majorana cones are protected by the combined $IT$ symmetry, where $I$ stands for inversion and $T$ the time reversal. Hence the Majorana cones remains robust in the spinon spectrum even after Gutzwiller projection.



In Tab.~\ref{tab:SL} we list the eigenvalues of fidelity matrix $F$ with different system-size. The minimal/maximal eigenvalue increases/decreases with system size, indicating that the eigenvalues are approaching $(1,1,1,1)$ in the large-size limit. This tendency suggests that the Gutzwiller projected state Z2Azz13SS is indeed $Z_2$ deconfined QSL. 
 The deconfinement of the $Z_2$ gauge field indicates the existence of 4-species of topological excitations, namely, $1, m, e, \varepsilon$, which correspond to the boson, vison ($Z_2$ flux), $Z_2$ charge and fermionic spinon, respectively. 

\begin{table}[htbp]
	\caption{Eigenvalues of the fidelity matrix in the SL phase with $J_1/J_2 = 0.71$, $J_r/J_2 = 0.6$. 
The minimal/maximal eigenvalue increases/decreases with system size, indicating 4-fold degenerate ground states on a torus in the large-size limit.}
	\label{tab:SL}
	\centering
	\renewcommand{\arraystretch}{1.2} 
	\begin{tabular}{|c|cccc|}
		\hline
		8$\times$8 & 0.0562 & 0.2249 & 1.2208 & 2.4980 \\
		\hline
		12$\times$12 & 0.1577 & 0.3837 & 1.2881 & 2.1705 \\
		\hline
		16$\times$16 & 0.3174 & 0.4897 & 1.2413 & 1.9516 \\
		\hline
	\end{tabular}
\end{table}

\subsubsection{The EP phase}

{The EP phase} is also a gapless $Z_2$ deconfined phase and shows up with the decreasing of $J_r$. It differs from the $Z_2$ QSL phase only by the spontaneous breaking of the mirror or glide symmetry. The ansatz of the EP phase can be obtained from Z2Azz13SS by introducing nonzero value of $\lambda_z$ and adding two extra parameters -- 
the bond modulation parameters $\eta_{x},\eta_{y}$ with $\eta_{x}\cdot \eta_{y}=1$. 
The resulting state explicitly breaks the mirror symmetry but preserve the $C_4$ lattice rotation symmetry (see Fig.\ref{fig:plaquette} and App.\ref{app:mf} for illustration). 
Since the $IT$ symmetry is also preserved, the spinon spectrum is still gapless, but the positions of the four cones deviate from the diagonal $k_x\pm k_y$ lines due to the breaking of mirror symmetry.

\begin{table}[htbp]
	\caption{ Eigenvalues of the fidelity matrix for the EP phase with $J_1/J_2 = 0.7, J_r/J_2=0$. The data support 4-fold degenerate ground state on a torus.}
	\label{tab:EP}
	\centering
	\renewcommand{\arraystretch}{1.2} 
	\begin{tabular}{|c|cccc|}
		\hline
		8$\times$8 &0.3234 &  0.3410 &  1.2811 &  2.0546 \\
		\hline
		12$\times$12    & 0.9320 & 0.9834 & 1.0189 & 1.0658\\
		\hline
 		16$\times$16    & 0.9447 & 0.9772 & 1.0079 & 1.0703 \\
		\hline
	\end{tabular}
\end{table}

After Gutzwiller projection, the EP phase is $Z_2$ deconfined because the eigenvalues of the fidelity matrix (as shown in Tab.\ref{tab:EP}) still show 4-fold degeneracy on a torus. Furthermore, the maximum eigenvalue of the fidelity matrix is close to 1 compared to the QSL phase, inferring that the the $Z_2$ flux excitation gap (or vison gap) is even larger than that in the QSL phase.

Actually, even without $\eta_{x,y}$, the $\lambda_z\tau_z$ term already breaks the PSG mirror reflection symmetry because $\lambda_z$ reverse its sign under the PSG operation
$$
 \tilde M_{x + y} = M_{x + y} (-1)^{i_x+i_y} \exp\{-i{\tau_x\over2}\pi\},
$$
where $M_{x + y}$ stands for the pure lattice mirror reflection. 
Our numerical calculations indicate that $\Delta_2 \neq 0$ and $t_2 \approx 0$ in the EP phase, and that the sign of $\lambda_z \cdot \Delta_2$ determines the pattern of the EP order (see Fig.\ref{fig:plaquette}(a)), namely, if the sign of $\lambda_z$ or $\Delta_2$ is reversed, then the EP order will be shifted to the alternative pattern. More details for the EP phase can be found in App.~\ref{app:mean field rules}. 


\subsubsection{The FP phase}

\begin{figure}[t]
	\centering	
\includegraphics[width=1\linewidth]{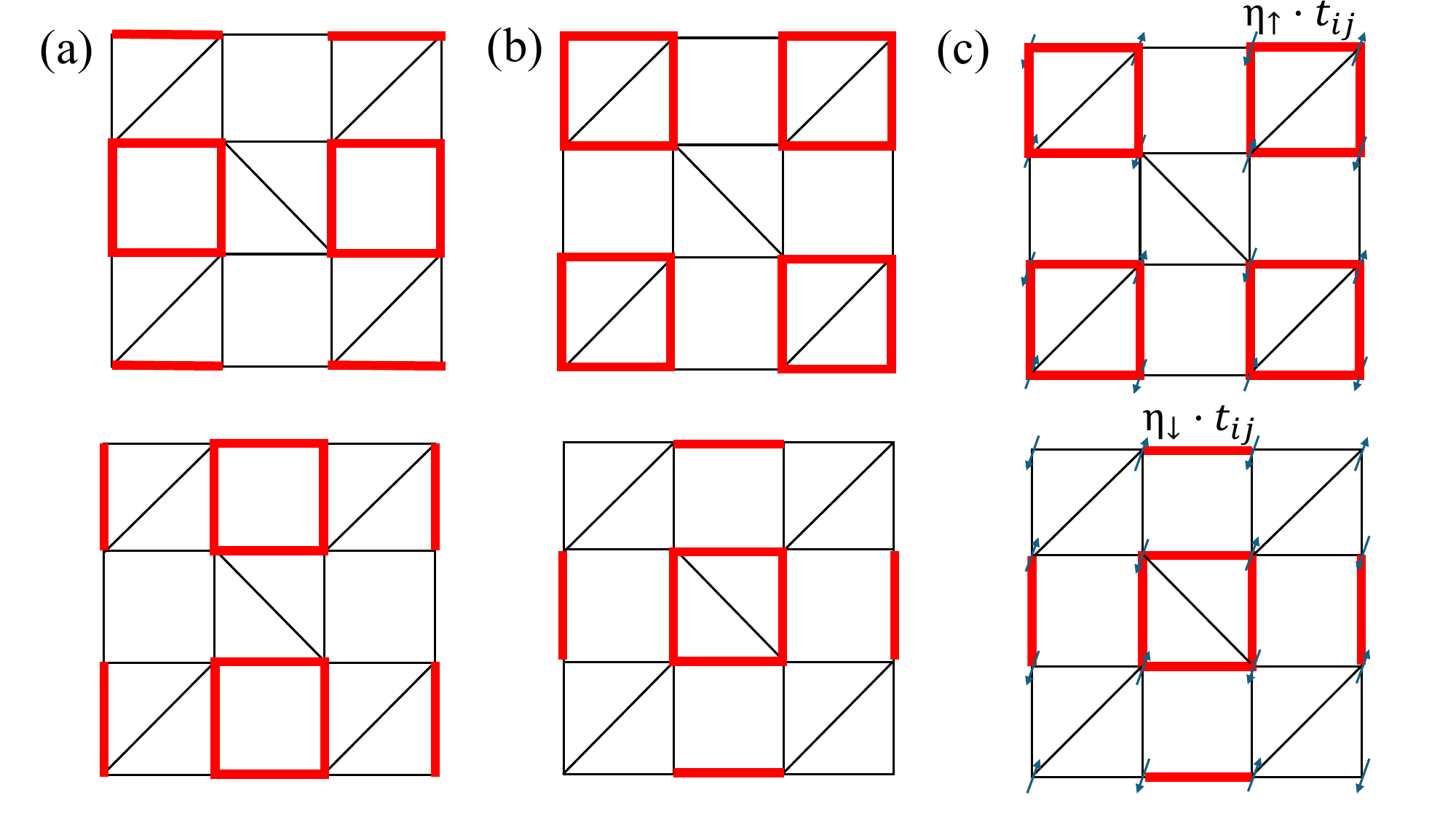}
	\caption{
		(a)Two patterns of empty plaquette order in the EP phase (uper pattern: $\eta_x>1, \eta_y<1$; lower pattern: $\eta_x<1, \eta_y>1$).
		(b) Two patterns of full plaquette order in the FP phase (uper pattern: $\eta_x>1, \eta_y>1$; lower pattern: $\eta_x<1, \eta_y<1$).
		(c) Ansatz of the N\'eel phase having SSG symmetry, where the upper/lower graph respectively represents the hopping of the $f_\up$/$f_\dn$ spinons, and the red lines indicate enhanced hopping amplitudes.
	}
	\label{fig:plaquette}
\end{figure}

{The FP phase} appears in the large $J_r$ region, in which the strong bonds are located on the plaquettes with diagonal links. The FP order is characterized by $\eta_x, \eta_y>1$ (or $\eta_x, \eta_y<1$) in the mean field Hamiltonian, and the value of the chemical potential $\lambda_z$ is suppressed to be zero by the ring-exchange interaction. The mirror symmetry is now preserved, but the $C_4$ rotation symmetry breaks down to $C_2$ (see Fig.\ref{fig:plaquette}(b) for illustration).  

Although the spinon spectrum is still gapless, the $Z_2$ gauge field is suffering from confinement (eigenvalues of the fidelity matrix shown in Tab.\ref{tab:FP} infer the confinement in large size limit). With the spinons being confined, an elementary excitation is the combination of a pair of spinons which carry integer quantum numbers. 

\begin{table}[htbp]
	\caption{ The eigenvalues of the fidelity matrix for the FP phase with $J_r/J_2 = 0.9, J_1/J_2=0.62$. The max eigenvalue of the fidelity matrix increase with size, indicating the tendency of single ground state on a torus and $Z_2$ confinement. }	\label{tab:FP}
	\centering
	\renewcommand{\arraystretch}{1.2}
		\centering
		\begin{tabular}{|c|cccc|}
			\hline
			8$\times$8  &  0.1037 &  0.2522 &  1.2822&   2.3618 \\
			\hline
			12$\times$12  & 0.0697  & 0.2411  & 1.0830  & 2.6062  \\
			\hline
			16$\times$16   &0.0328  & 0.165 &   0.9927 &  2.8087	\\			
			\hline
		\end{tabular}
\end{table}

\subsubsection{The N\'eel phase}

{The N\'eel phase} appears when $J_1$ is large. The optimal variational parameters favor finite magnetization $M_z$,  nonzero $t_2$, and vanishing $\Delta_2, \lambda_z$. Furthermore, the $f_\up$ fermions have larger hopping amplitude surrounding the full plaquettes with right-up diagonal bonds, while the $f_\dn$ fermions have larger hopping amplitude surrounding the rest full plaquettes (see Fig.\ref{fig:plaquette}(c) for illustration). This splitting structure is energetically robust when both $t_2$ and $M_z$ are nonzero. While the mean field Hamiltonian seems to break the lattice $C_4$ rotation symmetry, it is invariant under the composition of $C_4$ and spin flipping, namely $(C_{2x}|| C_{4})$, here we assume that the N\'eel order is parallel to $z$-axis. Similar symmetries include $(E||M_{x+ y})$ and $(C_{2x}||M_y|t_y)$.  
These operations are typical spin space group\cite{LiuQiHangPRX24, SongZhidaPRX24, YangFangLiuPRX24} elements where the spin rotation is unlocked with the lattice rotation. As a result, the spinon energy spectrum has an altermagnetic spin-splitting structure. 

\begin{table}[b]
	\caption{Eigenvalues of the fidelity matrix for the N\'eel phase with $J_r/J_2 = 0, J_1/J_2=0.74$. The max eigenvalue of the fidelity matrix increase with size, indicating the tendency of single ground state on a torus and $Z_2$ confinement.}
	\label{tab:AFM}
	\centering
	\renewcommand{\arraystretch}{1.2}
	\begin{tabular}{|c|cccc|}
		\hline
        8$\times$8 & 0.0740 &  0.1650 &  1.1012 &  2.6598 \\
        \hline
		12$\times$12 & 0.0216 & 0.0674 & 0.9986 & 2.9120 \\
		\hline
		16$\times$16 & 0.0139 & 0.0303 & 0.9994 & 2.9564 \\
		\hline
	\end{tabular}
\end{table}

Similar to the EP phase, the sign of $t_2\cdot M_z$ determines the pattern of amplitude modulation of the hopping terms of the spinons, namely, reversing the sign of $t_2\cdot M_z$ will shift the pattern of the strong-weak pattern of the hopping amplitudes of the $f_\up$  and $f_\dn$ fermions.

The $Z_2$ gauge field is confined in the N\'eel phase according to the eigenvalues of the fidelity matrix shown in Tab.~\ref{tab:AFM}.

The spinon mean-field spectrum remains gapless, but due to the confinement of the spinons, the low energy excitations are gapless magnons instead of spinons. The spin space group symmetry also affects the magnon band structure\cite{Chen2025Magnons,songLiu2024}.

\section{ Phase transitions} \label{sec:PT}

\begin{figure}[t]
	\centering
	\	\includegraphics[width=0.85\linewidth]{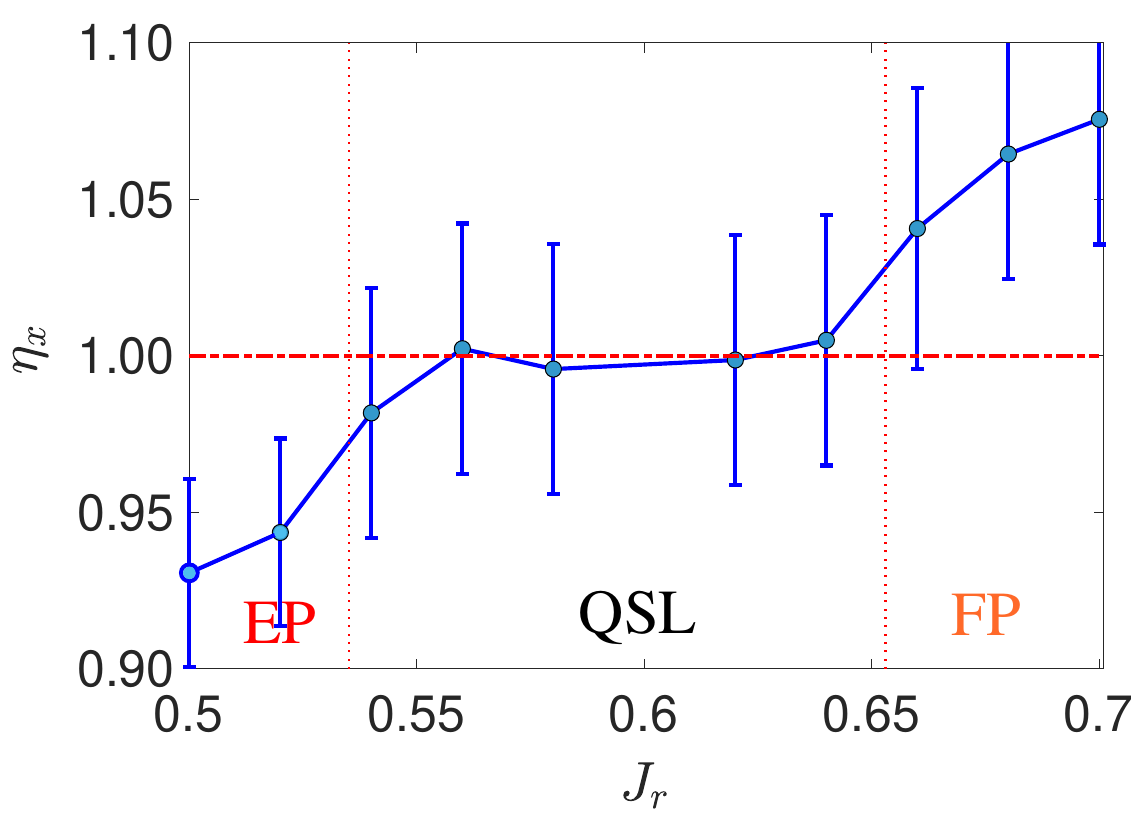}
	\caption{Evolution of the plaquette order parameter $\eta_x$ from the QSL phase to the EP and FP phases ($J_1/J_2=0.71$). In the QSL phase, $\eta_x=\eta_y\simeq 1$ indicating the absence of plaquette order; in the EP phase, $\eta_x<1, \eta_x\cdot \eta_y=1$; in the FP phase, $\eta_x=\eta_y>1$.}
	\label{fig:TransQSLEPFP}
\end{figure}

In this section, we analyze the phase transitions between different phases. The solid lines in the phase diagram Fig.~\ref{fig:phasedgrm} represent first-order transitions due to the discontinuity of the variational parameters. The dotted lines separating the QSL phase with neighboring phases stand for continuous phase transitions. 
Fig.\ref{fig:TransQSLEPFP} illustrate the evolution of the plaquette order parameter $\eta_x$ from the QSL phase to the EP and FP phases. In the following, we try to unveil the nature of these continuous transitions. More general description illustrated in App.\ref{app:Phasetransitions}.

\subsubsection{QSL to EP}
The phase transition from QSL to EP is relatively simple. According to our VMC study, the EP and the QSL phases are both $Z_2$ deconfined, the transition between them is characterized by the spontaneous breaking of the mirror reflection symmetries $M_{x\pm y}$ (as well as $G_x$ and $G_y$). 

If we consider the mirror reflection as `Ising symmetry', and treat the plaquette order parameters $\eta_{x,y}$ as the Ising order parameter, then the phase transition is in analog to the Ginzburg–Landau type transition describing the spontaneous Ising symmetry breaking. The continuous onsetting of the plaquette `order parameter' $\eta_x$ when entering the EP phase from the QSL side is shown in Fig.\ref{fig:TransQSLEPFP}. According to Ginzburg–Landau theory, the spontaneous breaking of the mirror reflection symmetry indicates that there are gapless Bosonic modes at the critical point, accompanying with divergence of the correlation length of the plaquette order (as well as the domain walls) and the corresponding generalized susceptibility.  

The subtlety is that the gapless Bosonic modes 
may have nontrivial coupling with the gapless fermionic spinons 
recalling that the pattern of the plaquette order is related to the sign of $\lambda_z \cdot \Delta_2$. Since $\lambda_z$ and $\Delta_2$ terms are both fermion bilinear, the coupling between the Ising order and the fermions, if it exists, is inferred to be biquadratic in the fermion operators.  

\subsubsection{QSL to the confined FP and N\'eel phases}

The transition from a gapped $Z_2$ spin liquid   to confined phases can be understood in terms of anyon condensation. For instance, the condensation of $m$ particle (the $Z_2$ flux, also called vison) at a nonzero momentum point gives rise to a valence bond solid order\cite{XucenkeBalents_vison, Xucenkesachdev_vison}; while the condensation of the $e$ particle (the composite of fermionic spinon and $m$ particle) results in long-range magnetic order \cite{Wangfa_schwingerboson, Wang2006SchwingerBoson}. However, situations are different for gapless systems, where the scenario of anyon condensation is no longer suitable because the gapless fermions may participate in the low-energy dynamics. For instance, on the square lattice systems the mechanism for the transition from a gapless $Z_2$ QSL to a confined  phases had been proposed\cite{Sachdev_SL, Senthil_Z2toAFM, PNAS_Z2toAFM}. The key idea is that the $Z_2$ gauge field is enhanced to $U(1)$ or $SU(2)$ at the critical point (previous VMC calculation~\cite{VMC_delta2} indicates that the pairing term $\Delta_2$ reduces to zero). Then the proliferation of monopoles results in $U(1)$ confinement\cite{Polyakov_book} and drives the system into the N\'eel or the VBS phase\cite{Song2019, Song2020}, depending on the microscopic spin-spin interactions.

However, situations are not the same in the SS lattice model (\ref{J1J2J4}). The key difference is that the fermion pairing $\Delta_2$ remains finite at the transition point,  hence the IGG is $Z_2$ throughout the phase transition. Other differences include: the positions of the fermionic nodes deviate from the high symmetry point $(\pi,\pi)$ by an offset $\big(\pm \tfrac{1}{2} \tfrac{\sqrt{\Delta_2^2+t_2^2}}{\Delta_1},\pm \tfrac{1}{2} \tfrac{\sqrt{\Delta_2^2+t_2^2}}{\Delta_1}\big)$;
the N\'eel phase has a spin splitting structure. The mechanism of such $Z_2$ confinement transitions is challenging. 

We conjecture that the $Z_2$ vison gap (of the Gutzwiller projected state) closes at the transition points even if the pairing term $\Delta_2$ is nonzero. The condensation of the gapless $m$ particle (or $e$ particle) together with the fermion bilinears result in a $Z_2$ confinement phase with VBS (or magnetic) order. This possibly interprets the QSL-FP phase transition. But the magnetic order from this $Z_2$ confinement transition should be non-collinear\cite{xucenke_Z2SLtoNeel, xucenke_Z2SLtoorder, donghai_Z2SLtoorder}, which is inconsistent with the N\'eel order. There are several possible solutions: (i) the ring-exchange interaction may drive the magnetic order slightly away from the collinear one and give rise to an incommensurate order; (ii) the magnetic order is still N\'eel order, but after the transition there is an extra AFM$^*$ phase having emergent $Z_2$ gauge field and coexisting magnetic order; (iii) the critical point is indeed $Z_2$ confinement transition to a N\'eel phase, which is a completely new mechanism; (iv) the confinement transition is of weak first order.

\subsubsection{Possible criticality between EP and N\'eel}

We finally discuss the possible criticality between EP and N\'eel phase. According to our VMC computation, the transition from EP to N\'eel phase is first-order because the N\'eel order and other parameters are discontinuous at the transition line. 
While the symmetry breaking orders (like the N\'eel order) may be over estimated\cite{Becca_VMC}, 
the chemical potential $\lambda_z$ also exhibits a clear jump 
on the phase boundary when $J_r \lesssim 0.2$ (see App.~\ref{app:Phasetransitions} and Fig.~\ref{fig:J4}). 
However, when $J_r \gtrsim 0.2$, the discontinuity is suppressed and the transition between the two phases is close to a continuous one, which is forbidden by the Ginzburg-Landau paradigm. 

In literature, a continuous transition betweem a N\'eel phase and a gapped nematic spin liquid was proposed in Ref.\onlinecite{Qiyang_NeeltoZ2} via condensation of spinon pair-skyrmion bound states. Later a confinement transition from a gapless $Z_2$ QSL to a N\'eel phase was studied on the square lattice where the Ising flux susceptibility diverges at the critical point\cite{PNAS_Z2toAFM}, indicating the closure of the vison gap. However, a continuous transition from a gapless $Z_2$ deconfined EP phase to the N\'eel phase was never studied before. 

For the occurrence of such a transition, the following conditions should be satisfied simultaneously: (i) confinement the $Z_2$ gauge field; (ii) spontaneous breaking of the $SU(2)$ spin-rotation symmetry; and (iii) restoration of the mirror symmetry. 


As mentioned, reversing the sign of $\lambda_z \Delta_2$ will switch the pattern of the EP order in the EP phase. To create a pair of $m$ particles one need to reverse the sign of $\Delta_2$ on the bonds crossing the vison line (in analog to the Dirac string), given that the sign of $\lambda_z$ remains unchanged. Therefore, creating $m$ particles will give rise to local shifting of the pattern of the EP order in a certain region. In other words, due to the sign rule observed from VMC, the $m$ particle is associated with certain kind of defect of the EP order. Therefore, if the gap of the $m$ particles closes at the transition to the confined phase, the critical point will be accompanied by the simultaneous condensation of the defects of the EP order, which leads to the restoration of mirror symmetry.

 In the above discussion, the $\Delta_2$ term, i.e. the pairing on the diagonal bonds, which plays the role of Higgs field, remains finite throughout the phase diagram. The chemical potential $\lambda_z$ is another Higgs field which vanishes outside the EP phase.
Therefore, the critical point between EP and N\'eel phases, if exist, is a $Z_2$ confinement transition. This criticality is very different from the DQCP in square lattice proposed in literature \cite{Tanaka-Hu_piflux,YouPRX,spectral_signature_piflux} where the critical theory is described by a $\pi$-flux ansatz.

\section{ Conclusion and discussion}\label{sec:conc}

In summary, we studied the $J_1$-$J_2$-$J_r$ model on the Shastry-Sutherland lattice via VMC method.  Besides the dimer phase, the antiferromagnetic ordered N\'eel phase, the empty plaquette (EP) phase (which spontaneously break the mirror reflection symmetry) and the full plaquette (FP) phase (which spontaneously break the $C_4$ rotation symmetry), a gapless $Z_2$ QSL phase is observed which is closely related to the Z2Azz13 gapless $Z_2$ QSL on the square lattice. Different from literature, the $EP$ phase in our VMC appraoch is a $Z_2$ deconfined phase with massless majoran-type spinon excitations. While the phase transitions between two ordered phases are generally of first order, the transition between the QSL and the ordered phases are continuous. The criticality between the QSL and the EP phases is characterized by spontaneous mirror symmetry breaking and may be described by the Ginzburg-Landau scenario. What is exotic is that the transitions between the QSL and the N\'eel and FP phases are $Z_2$ confinement transitions where the Higgs field is always condensed throughout the transitions, hence the gauge fluctuations are always of $Z_2$ type. 

Since the gapless spinon excitations form Majorana cones, the EP and QSL phases can be potentially detected in experiments via power law temperature dependence of the magnetic specific heat $C_v (T) \sim T^2$ or the magnetic susceptibility $\chi(T)\sim T$  at low temperatures. 

We proposed possible interpretations of these transitions but their final mechanism need further investigation. Furthermore, we conclude that the critical point between the gapless EP phase and the N\'eel phase, if exist, is also a $Z_2$ confinement transition, which is different from the DQCP proposed in literature. 
 
When finalizing the present work, we noticed a related work \cite{VMC_SL} which classified QSLs on the SS lattice. Our QSL phase is consistent with the classification. 

{\it Acknowledgement.} We thank Yang Qi, Wei Zhu, Rong Yu, Zhi-Yuan Xie and Yi-Zhuang You for helpful discussions. This work was supported by National Basic Research and Development plan of China (Grants No.2023YFA1406500, 2022YFA1405300) and NSFC (Grants No.12134020, 12374166). Computational resources have been provided by the Physical Laboratory of High Performance Computing at Renmin University of China.

\bibliographystyle{apsrev4-2}
\bibliography{references}

\newpage
\appendix
\widetext 

\section{$J_1$-$J_2$ model} \label{app:J1J2}
As $J_r=0$, the EP phase is observed in the region $0.698<J_1/J_2<0.719$, which is sandwiched by the dimer phase at $J_1/J_2<0.698$ and the N\'eel phase at $J_1/J_2>0.719$. Theses phases is consistent with the DMRG study\cite{YouPRX}. However, different from the gapped phase proposed in SBMFT\cite{Wangfa}, the EP phase in our VMC approach is a $Z_2$ deconfined phase with gapless spinon excitations. Furthermore, the N\'eel phase is characterized by a spin space group (SSG) symmetry  \cite{LiuQiHangPRX24, SongZhidaPRX24, YangFangLiuPRX24} with spin splitting in the excitation spectrum. Our results reveal a first-order phase transition between the EP phase and N\'eel phase. 

\begin{figure}[htbp]
	\centering
	\includegraphics[width=.65\linewidth]{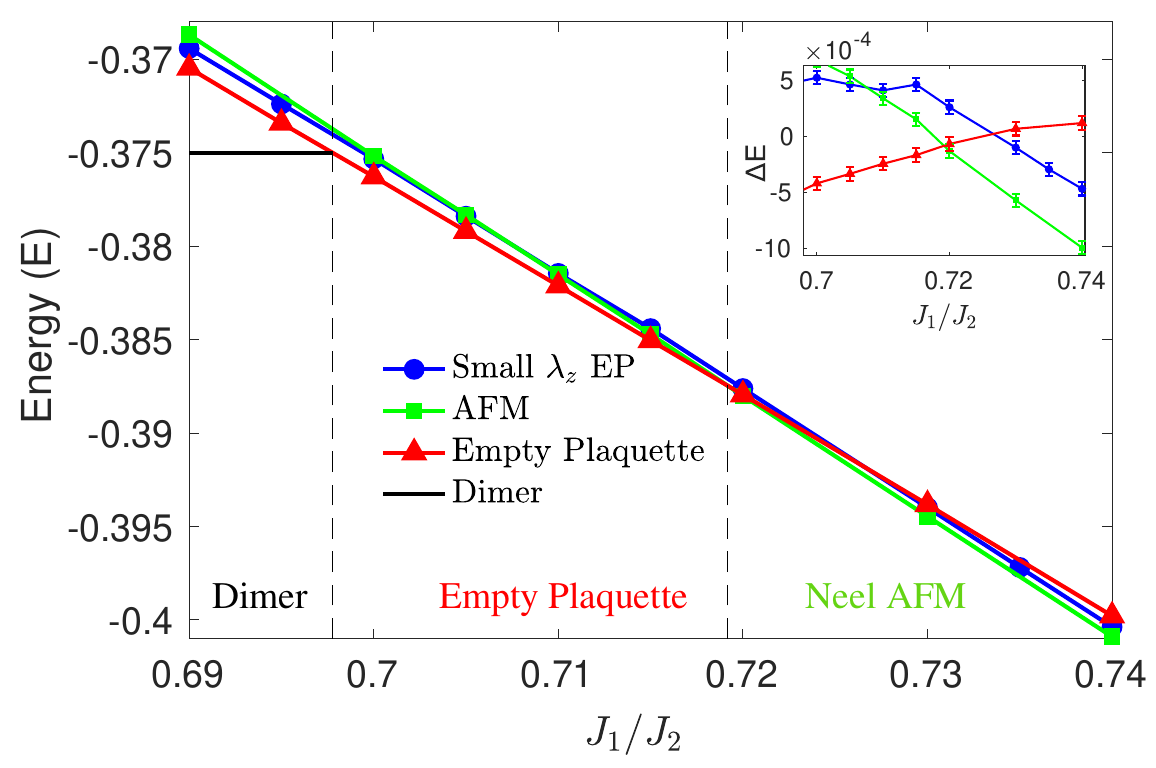}
	\caption{Phase diagram of \( J_1$-$J_2 \) SS model. The region \( J_1/J_2 < 0.698 \) is the dimer phase and the intermediate region \( 0.698 < J_1/J_2 < 0.719 \) corresponds to a \( Z_2 \) deconfined EP phase, and the right region \( J_1/J_2 > 0.719 \) stands for the N\'eel phase.  
		The inset in the top-right corner shows the energy difference relative to a fitted baseline. All transitions are of first order.}
	\label{fig:J1J2phase diagram}
\end{figure}

\section{Relation between the QSL in SS and the Z2Azz13 on the Square lattice}\label{app:Z2Azz13}

If the mean field parameters preserve all the symmetries of certain PSG, then the resulting projected state is a QSL. It is known that for the $J_1$-$J_2$ Heisenberg model on the square lattice, the energetically favored ansatz is the Z2Azz13 gapless $Z_2$ QSL~\cite{ HuwenjunSL,Sachdev_SL,wen_PSG}. In our study of the SS model, after numerous trials, we find that the possible lowest-energy ansatz also descending from the Z2Azz13 type, which is consistent with the conclusions of the similar work\cite{VMC_SL}. This is quite reasonable, considering that the SS lattice can be viewed as a square lattice with the diagonal ($J_2$) bonds partially removed. Therefore, it is natural that both lattices share the same type ansatz. Other higher energy ansatz are not the focus of our study, so we will not draw them in the phase diagram. Then the mean field ansatz of symmetry breaking phases are constructed based on this QSL ansatz by adding symmetry breaking parameters (see App.\ref{app:mf}).

\subsection{Three different gauge choices}

Numerically, we usually adopt the $d$-wave representation. Before introducing the symmetry-breaking orders, in QSL phase, we employ the following ansatz:
\begin{equation}\label{uij}
	\begin{aligned}
		& {u}_{i, i+\hat{x}}=\left(\begin{array}{cc}
			t_1 & \Delta_1 \\
			\Delta_1^* & -t_1^*
		\end{array}\right),\quad
		{u}_{i, i+\hat{y}}=\left(\begin{array}{cc}
			t_1 & -\Delta_1\\
			-\Delta_1^* & -t_1^*
		\end{array}\right), 
		\\
		& {u}_{i, i+\hat{x}+\hat{y}}=\left(\begin{array}{cc}
			t_2 & \Delta_2 \\
			\Delta_2^* & -t_2^*
		\end{array}\right),\quad
		{u}_{i, i-\hat{x}+\hat{y}}=\left(\begin{array}{cc}
			t_2 & -\Delta_2 \\
			-\Delta_2^* & -t_2^*
		\end{array}\right),\\
		&
		\quad
		{u}_{i, i}=\left(\begin{array}{cc}
			\lambda_z &0\\
			0 & -\lambda_z
		\end{array}\right), \quad
		{u}^{(3)}_{i, i}=\left(\begin{array}{cc}
			M_z & 0 \\
			0 & M_z
		\end{array}\right), 
	\end{aligned}
\end{equation} 
we can fix $t_1=1$ in VMC calculations.  $u^{(3)}_{i,i}$ is the third component of on-site background field of long-ranged magnetic order. If chemical potential and N\'eel order is zero, then this ansatz can be obtained by deforming
the Z2Azz13 ansatz on the square lattice 
to the SS lattice by removing certain diagonal $J_2$ bonds followed by sequence of gauge transformations. Noticing that there are two typical gauge choices for the Z2Azz13 --- dubbed stagger-flux gauge and pure-paring gauge, in the following we perform gauge transformations to identify the above ansatz (\ref{uij}) with the stagger-flux  ansatz and pure-pairing ansatz. 
 
We first consider the stagger-flux gauge choice\cite{Sachdev_SL}. By performing a gauge transformation $W_i=\exp \left\{ i(-1)^{i_x+i_y} \frac{\pi}{4} \tau^x\right\}$, the above ansatz (\ref{uij}) with real hopping and real $d$-wave pairing terms can be transformed into the `stagger-flux' hopping terms on the $J_1$ bonds plus complex $d_{xy}$-wave pairing terms on the diagonal $J_2$ bonds. 
The resulting ansatz is analogous to the square-lattice model\cite{Sachdev_SL}:
\begin{equation}\label{tilde_uij}
	\begin{aligned}
		& \tilde{u}_{i, i+\hat{x}}=\left(\begin{array}{cc}
			t e^{-i \phi} & 0 \\
			0 & -t e^{i \phi}
		\end{array}\right), \quad i_x+i_y=\text{ even},\\
		&\tilde{u}_{i, i+\hat{x}}=\left(\begin{array}{cc}
			t e^{i \phi} & 0 \\
			0 & -t e^{-i \phi}
		\end{array}\right), \quad i_x+i_y=\text{ odd} \\
		& \tilde{u}_{i, i+\hat{y}}=\left(\begin{array}{cc}
			t e^{i \phi} & 0 \\
			0 & -t e^{-i \phi}
		\end{array}\right), \quad i_x+i_y=\text{ even},\\
		&\tilde{u}_{i, i+\hat{y}}=\left(\begin{array}{cc}
			t e^{-i \phi} & 0 \\
			0 & -t e^{i \phi}
		\end{array}\right), \quad i_x+i_y= \text{ odd} \\
		& \tilde{u}_{i, i+\hat{x}+\hat{y}}=\left(\begin{array}{cc}
			0 & -\left(\gamma_1-i \gamma_2\right) \\
			-\left(\gamma_1+i \gamma_2\right) & 0
		\end{array}\right)\\
		&\tilde{u}_{i, i-\hat{x}+\hat{y}}=\left(\begin{array}{cc}
			0 & \left(\gamma_1-i \gamma_2\right) \\
			\left(\gamma_1+i \gamma_2\right) & 0
		\end{array}\right)\\
		& \tilde{u}_{i, i}=\left(\begin{array}{cc}
			0 & -i\mu \\
			i\mu & 0
		\end{array}\right), \quad i_x+i_y= \text{ even}, \\
		&\tilde{u}_{i, i}=\left(\begin{array}{cc}
			0 & i\mu\\
			-i\mu & 0
		\end{array}\right), \quad i_x+i_y= \text{ odd},\\
		&\tilde{u}^{(3)}_{i, i}=\left(\begin{array}{cc}
			M_z & 0\\
			0 & M_z
		\end{array}\right).
	\end{aligned}
\end{equation}
with $ t=\sqrt{t_1^2+\Delta_1^2} ,\phi=\arctan \Delta_1/t_1, \gamma_1=-\Delta_2$, $\gamma_2=-t_2$, $\mu=\lambda_z$. In this gauge choice, the parameter $t_2$ can be transformed into $\Delta_2$ and vice versa by an uniform gauge rotation along $\tau^z$ direction if $\lambda_z$ is 0. We will use this relation in later discussion.

Then we consider the pure-pairing gauge of Z2Azz13 spin liquid, which was first proposed by X.-G. Wen\cite{wen_PSG} for the $J_1$-$J_2$ Heisenberg model on square lattice:
\beq
&&u_{i, i+\hat{x}}  =\chi \tau^x-\eta \tau^y, \label{Wen_Z2}\\
&&u_{i, i+\hat{y}}  =\chi \tau^x+\eta \tau^y, \notag\\
&&u_{i, i+\hat{x}+\hat{y}}  =-\gamma_1 \tau^x, \notag\\
&&u_{i, i-\hat{x}+\hat{y}}  =\gamma_1 \tau^x.\notag
\eeq
Since all the $u_{ij}$ matrices are off-diagonal, 
the corresponding mean field Hamiltonian only contain pure-paring terms. 

\begin{figure}[t] 
    \centering
    \begin{subfigure}{0.4\linewidth}  
    \includegraphics[width=0.55\linewidth]{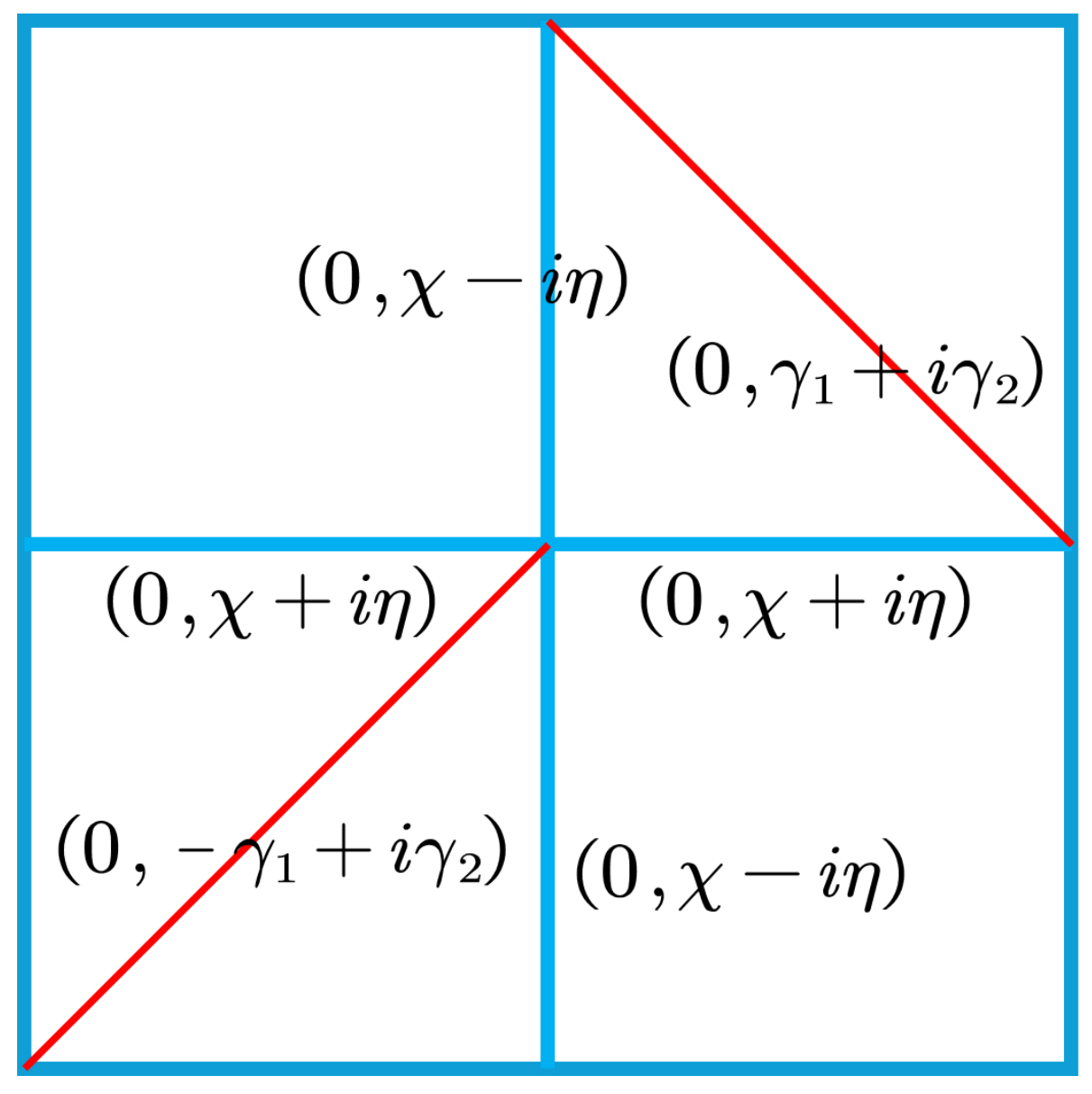} 
    \caption{The ansatz descendent from Z2Azz13 in the pure-pairing gauge}\label{Z2_1}
    \end{subfigure}
    \centering    
    \begin{subfigure}{0.4\linewidth}    
    \includegraphics[width=0.55\linewidth]{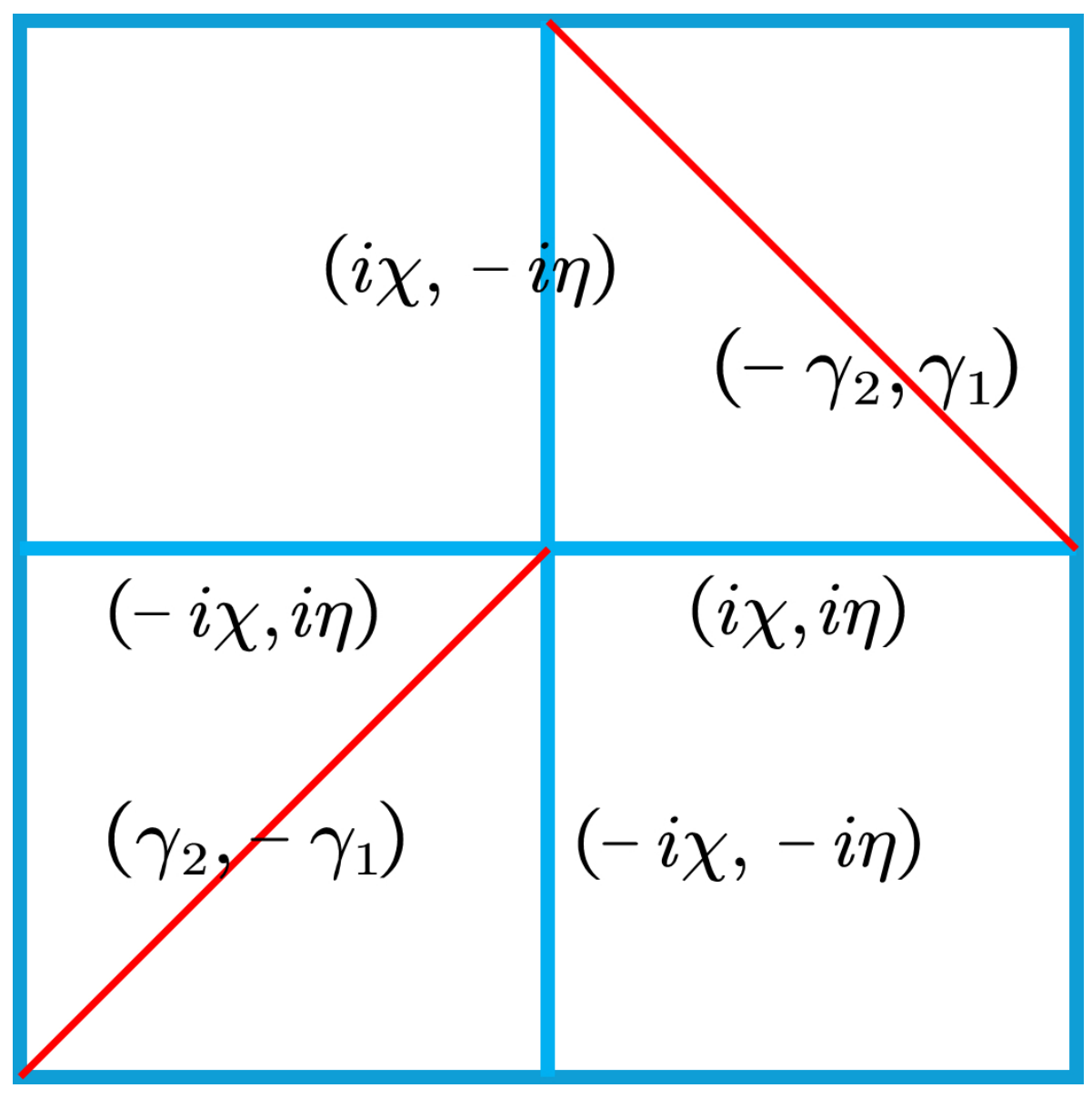}
    \caption{Acting gauge transform $W_i=\exp \left(i(-1)^{i_x+i_y} \frac{\pi}{4} \tau_1\right)$.}\label{Z2_2}    
    \end{subfigure}    
    \centering    
    \begin{subfigure}{0.4\linewidth}      
    \includegraphics[width=0.55\linewidth]{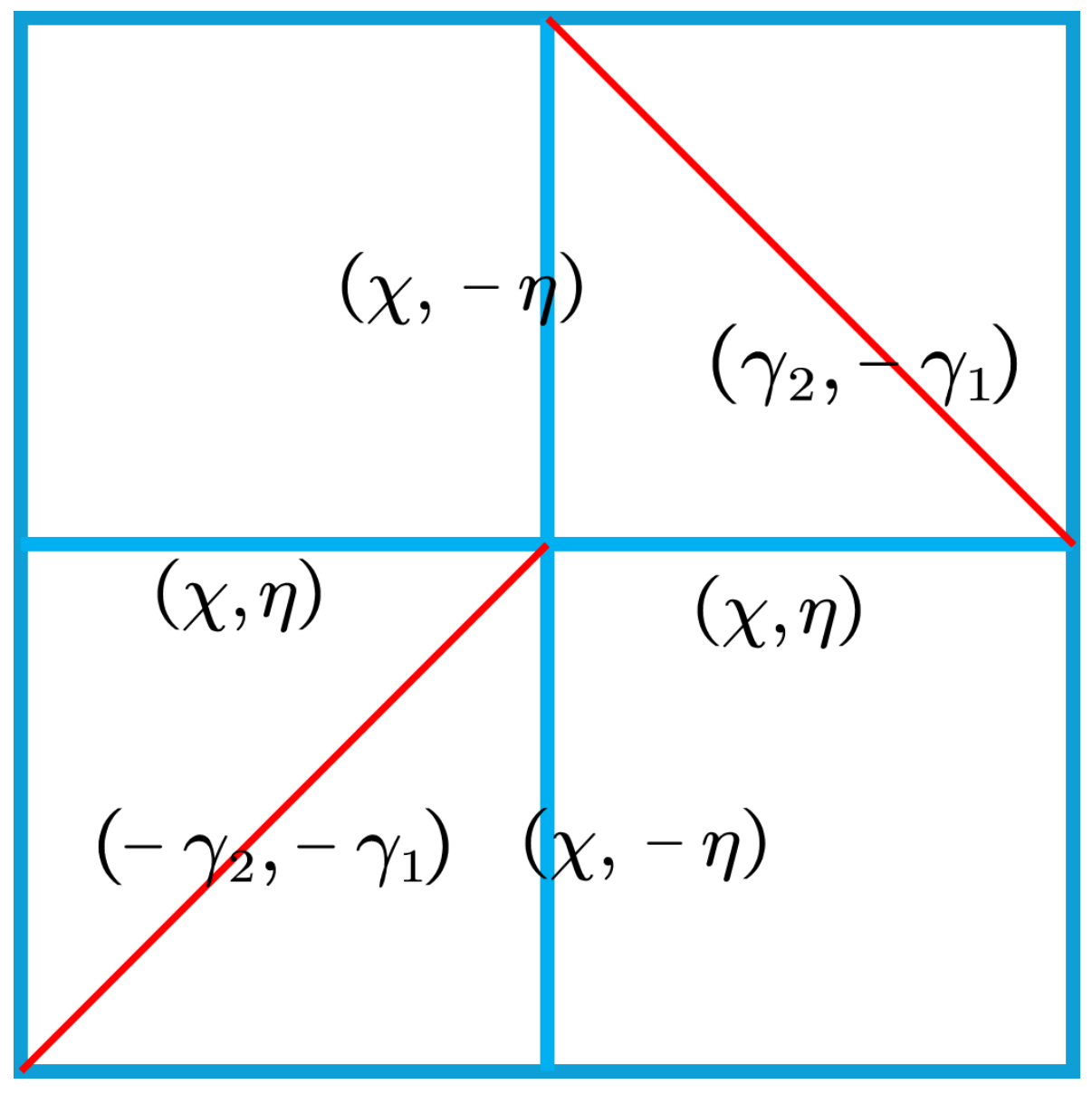}
    \caption{Acting gauge transformation  $W_{i{\rm e}}=\exp \left(i \frac{\pi}{2} \tau_1\right), W_{i{\rm o}}=\exp \left(-i \frac{\pi}{2} \tau_2\right)$.}\label{Z2_3}      
    \end{subfigure}
    \centering    
    \begin{subfigure}{0.4\linewidth}      
    	\includegraphics[width=0.55\linewidth]{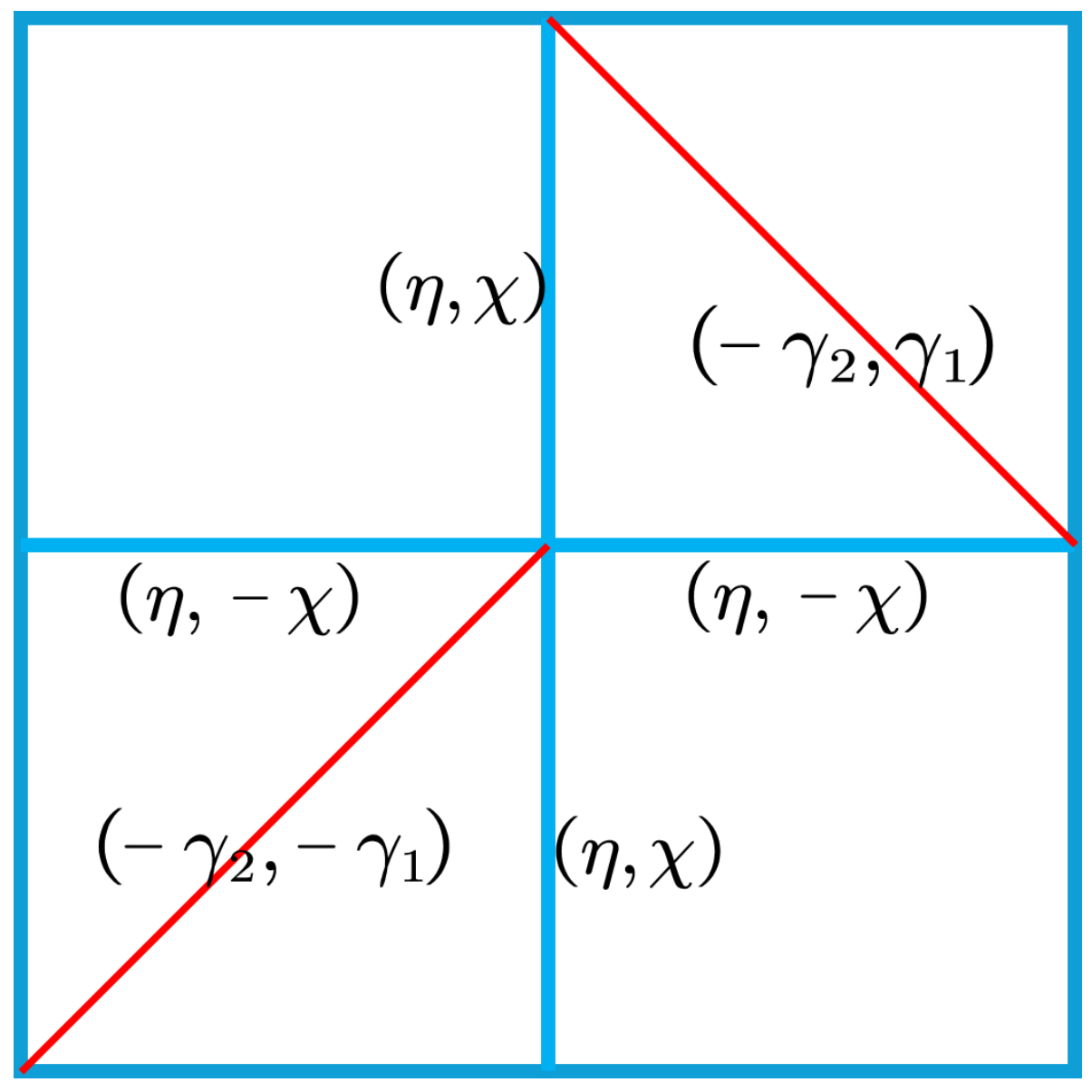}
    	\caption{Acting gauge transformation  $W_i=\{\,-i\tau_2,\;-1,\;i\tau_2,\;1\,\}$.}\label{Z2_4}      
    \end{subfigure}
	\caption{After the sequence of gauge transformations, the ansatz descended from Z2Azz13
	 on square lattice is turned into the ansatz Z2Azz13SS we discussed in the main text [see Fig.\ref{fig:SSmodel}(b) and (\ref{uij})]. }
	 \label{Z2}
\end{figure}

By removing certain diagonal bonds from the $J_1$-$J_2$ model on square lattice, one obtains the following ansatz on the SS lattice (see Fig.\ref{Z2_1}), where the first term in the bracket means the hopping term and the second term stands for the pairing. 



After series gauge transformations, one obtains the ansatz as shown in Fig.\ref{Z2}(d)-- the $Z_2$ QSL ansatz in the main text, where $W_{i \rm{e}},W_{i\rm{o}}$ respectively means the gauge transformation acting on even/odd site. In the final gauge transformation, the four matrix elements correspond to the clockwise action on the four sublattice sites within a unit cell. Setting $\chi=-\Delta_1,\eta=t_1$ and $\gamma_1=-\Delta_2$, $\gamma_2=-t_2$, one can identify the ansatz in Fig.\ref{Z2}(d) with Z2Azz13SS [see Fig.\ref{fig:SSmodel}(b)] of the main text. 

\subsection{PSG symmetry of Z2Azz13SS}\label{app:PSG}

Next, we will analyze the PSG of the SS model. The symmetry in SS lattice:
\Beq
		&&T_x(x,y) \to (x+2,\,y),\\
		&&T_y(x,y) \to (x,\,y+2),\\
		&&G_x(x,y) \to (x+1,\,-y),\\
		&&G_y(x,y) \to (-x,\,y+1),\\
		&&M_{x+y}(x,y) \to (y,\,x),\\
		&&C_4 (x,y) \to (-y+2,\,x-1).
\Eeq
The symmetry of the SS lattice can be generated from the symmetry operations of the square lattice. For example, combining two operations $O_2,O_1$ , the corresponding PSG transformation:
\begin{equation}
	G_{O_2}\!\big(O_2 O_1(i)\big)\,
	G_{O_1}\!\big(O_2 O_1(i)\big)\,
	u_{O_2 O_1(i),\,O_2 O_1(j)}\,
	G_{O_1}^{\dagger}\!\big(O_2 O_1(j)\big)\,
	G_{O_2}^{\dagger}\!\big(O_2 O_1(j)\big)
	= u_{ij}.
\end{equation}
For the known transformations \( G_{O_2}(i'') \) and \( G_{O_1}(i') \) (where $i''=O_2O_1(i),i'=O_1(i)$),  
if we fix the coordinate system with respect to \( i'' \),  then we get the PSG of the combining symmetry operation:
\begin{equation}
	G_{O_2 O_1}(i'') = G_{O_2}(i'') \, G_{O_1}\!\big(O_2^{-1}(i'')\big).
\end{equation}
From the operation relation: $T_x=t_x^2, G_x=p_xt_x, C_4= G_x\, \sigma_{xy}\, G_x\, G_y^{-1}=\sigma_{xy}p_y$. We obtain the PSG of SS lattice:
\begin{equation}
	\begin{aligned}
		W_{T_x}(i) &= \tau^{0}, &
		\qquad W_{G_x}(i) &= (-1)^{i_x+i_y}\, g_2(\theta)\,\mathrm{i}\tau^{y} , &
		\qquad W_{M_{x+y}}(i) &=(-1)^{i_x+i_y} g_2(\theta)\,\mathrm{i}\tau^{x}, \\
		W_{T_y}(i) &= \tau^{0}, &
		\qquad W_{G_y}(i) &= (-1)^{i_x+i_y+1}\, g_2(\theta)\,\mathrm{i}\tau^{y} , &
		\qquad W_{C_4}(i)&= \mathrm{i} \tau^z .
	\end{aligned}
\end{equation}

where $g_2(\theta)=e^{i (-1)^{i_x+i_y} \theta \tau_y},\theta=\arctan(t_2/\Delta_2)$. The above $g_2(\theta)$ originates from the $U(1)$ phase factor $g_3(\theta)$ in the staggered–flux (U1Cn01n ansatz) representation. Since in our representation we set $t_2$ and $\Delta_2$ to be nonzero,  a proper twist must be introduced when using the PSG to restore the diagonal bonds, leading to a judicious choice of the angle $\theta$. Finally, we can see that a nonzero chemical potential term $\lambda_z \tau_z$ explicitly breaks the mirror and glide symmetries while preserving the rotational symmetry, which is consistent with the symmetry requirements of the EP phase.

\section{Mean field ansatz and the corresponding phases}

\subsection{The variational parameters}\label{app:mf}

The general form of the mean field ansatz are given in Eq.(3) of the main text, and the meaning of the variational parameters are shown in Fig. \ref{fig:SSmodel}(b).
In the variational setup, the nearest neighbor hopping term \(t_1\) is normalized to be 1, and the next-nearest neighbor hopping \(t_2\) parameter and the pairing parameters \(\Delta_{1,2}\) are determined variationally. Here \(\Delta_1\) stands for \(d_{x^2-y^2}\) wave nearest neighbor pairing, and \(\Delta_2\) labels next-nearest neighbor \(d_{xy}\)-wave pairing term. 

\begin{figure}[h]
    \centering
    \includegraphics[width=0.6\linewidth]{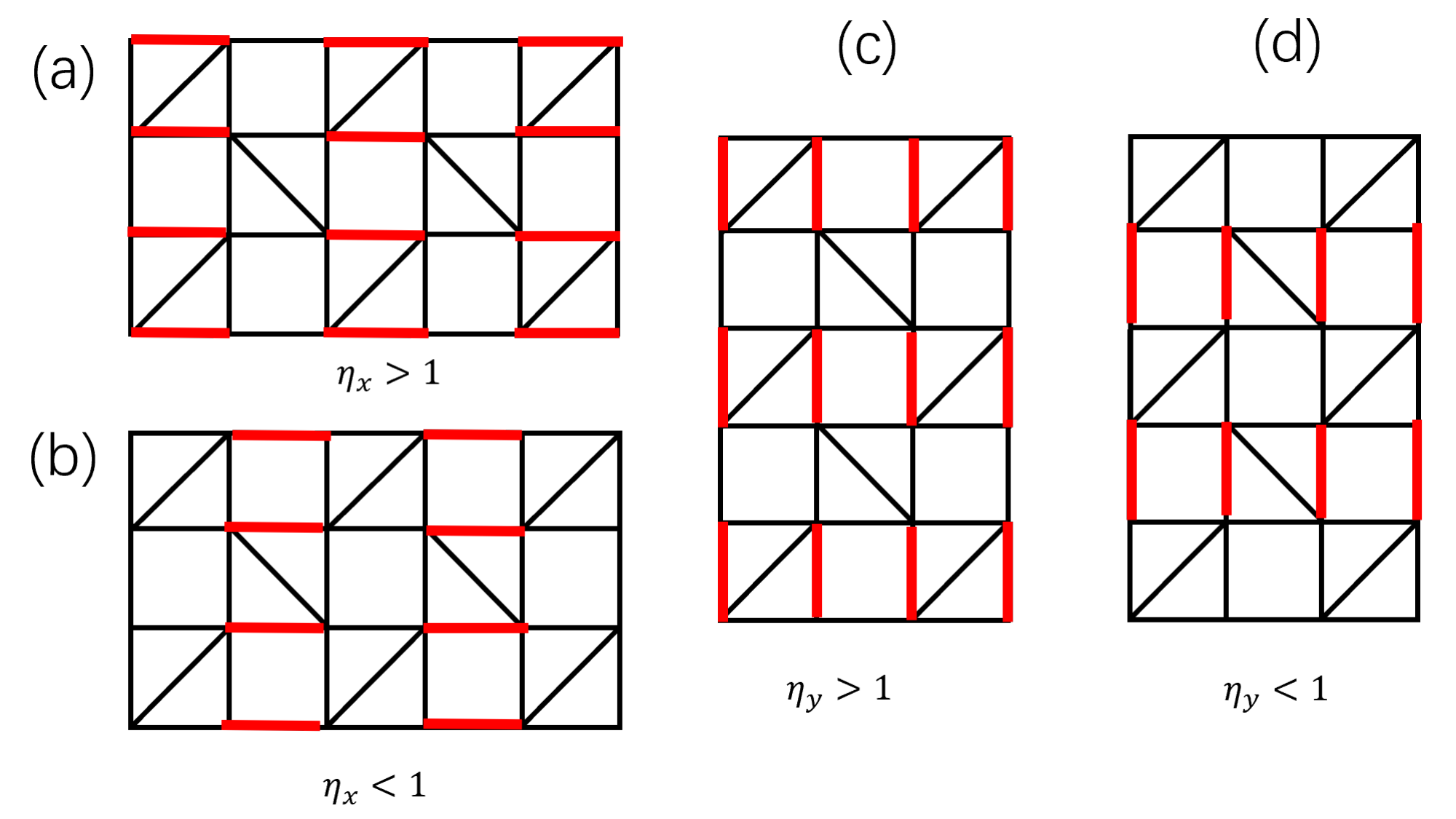}
    \caption{(a)(b) represent two types of VBS along the x direction, (c)(d) along the y direction.}
    \label{psvbs}
\end{figure}

In our variational study, the plquatte phases are parameterized by the staggered strong-weak pattern of the amplitudes of the variational parameters in Fig.\ref{psvbs}. Since the modulation of the amplitude of pairing terms has little effect on the trial energy, we only modulate the amplitudes of the hopping terms. There are two kinds of plaquette phases, namely, the EP phase and the FP phase. 

We include two parameter $\eta_x$ and $\eta_y$ in the variational Hamiltonian to encode the two kinds of plaquette-type valance bond solid (VBS) orders. As shown in Fig.\ref{psvbs}, $\eta_x$ and $\eta_y$ stand for the ratio of the hopping amplitude between the even-index bonds and odd-index bonds.
For simplicity, we normalize the hopping on {\it weak bonds} to be $t_1=1$, and set the hopping on the {\it strong bonds} as $\eta_{x,y} t_1$ (if $\eta_{x,y} >1)$ or $\frac{1}{\eta_{x,y}} t_1$ (if $\eta_{x,y} <1)$.
The case $\eta_x=\eta_y=1$ means no plaquette order. As illustrated in Fig. \ref{fig:plaquette} and Fig.\ref{psvbs}, the  parameters setting $\eta_x>1, \eta_y<1$ and $\eta_x<1, \eta_y>1$ stands for the two different patterns of the EP order; while $\eta_x>1, \eta_y>1$ and $\eta_x<1, \eta_y<1$ stands for the two different patterns of the FP order. Our VMC results automatically satisfy the relations $\eta_x\cdot \eta_y=1$ in the EP phase and $\eta_x=\eta_y$ in the FP phase. 

Through  the above analysis, variational parameters of the mean-field Hamiltonian includes:  $\Delta_1$,$t_2$,$\Delta_2$, hopping enlargement factor $\eta_x$ $\eta_y$, chemical potential $\lambda_z$, N\'eel order $(-1)^{x+y}M$. The ansatz in the EP phase is close to the QSL ansatz. Except for the mirror symmetry breaking parameter $\eta_{x,y}\neq 1$, a main difference is that the optimal value of the chemical potential ($i.e.$ the Lagrangian multiplier) term $\lambda_z \tau_z$ is zero in the QSL phase but nonzero in the EP phase. 

\subsection{ Identifying each phase } \label{app:mean field rules} 

We analyze the properties of each phase from their mean-field parameters which helps to understand the physics of the phase and phase transitions. This observation provides useful insight into the nature of the phase transition and inspires our analysis. 

\begin{table}[h]
	\centering
	\label{tab:parameters}
	\begin{tabular}{|c|c|c|c|c|c|c|c|c|}
		\hline
		 case  & E & $t_2$ & $\Delta_1$ & $\Delta_2$ & $\lambda_z$ & $\eta_x$ & $M_z$ & GSD (eigenvalues of $F$) \\ \hline
		(i) EP state  & -0.37627  & 0.0504 & 0.3669 & 0.5947  & 1.0402   & 0.7738 & -0.0072 & 0.9320, 0.9834, 1.0189, 1.0658 \\ \hline
		(ii) $\text{switch plaquette}$&  -0.36134  & 0.0504 & 0.3669 & 0.5947  & 1.0402   & 1.2957 & -0.0072 & 0.8151, 0.9904, 1.0261, 1.1684  \\ \hline
		(iii) $\text{switch plaquette,-$\Delta_2$}$ & -0.37621 & 0.0504 & 0.3669 & -0.5947  & 1.0402   &1.2957 & -0.0072 &   \\ \hline
		(iv) $\text{No plaquette order}$   & -0.370381 & 0.0504 & 0.3669 & 0.5947  & 1.0402   & 1 & -0.0072 & 0.9211, 0.9752, 1.0254, 1.0783 \\ \hline
		(v) $\text{other EP state}$ & -0.37622 & 0.2058 & 0.4378 & 0.7149 & 1.0571 & 0.7664 & 0.0144 & \\ \hline
		(vi) EP phase in $16\times16$ sites & -0.37642  & 0.0113 & 0.3966 &  0.5908  &  0.7736 &0.7761 &  0.0118 & 0.9447, 0.9772, 1.0079, 1.0703 \\ \hline
		(vii) $\text{no potential, $t_2$ dominate}$ & -0.37539 & 0.8227 & 0.7063 & 0.0174 & 0 & 1.2729 & 0.0050 & 0.3485, 0.5535, 1.0000, 2.0981 \\ \hline
		(viii) $\text{no potential, $\Delta_2$ dominate}$ & -0.37527 & 0.0794 & 0.5760 & 0.6387 & 0 & 1.1860 & -0.0641 & \\ \hline
		(ix) $\text{small potential}$ & -0.37537 & -0.0651 & 0.4498 & 0.5478 & -0.0828 & 1.3231 & 0.0081 & \\ \hline
		(x) $\text{no $\Delta_2$}$ & -0.37308 & 0.0464 & 0.8768 & 0 & 0.2774 & 1.0218 & -0.1542 & \\ \hline
		(xi) $\text{splitting N\'eel phase}$ & -0.37512 & 0.6423 & 0.6167 & 0.0639 & 0 & 1.4036$(\eta_{\uparrow})$& 0.2017  & \\ \hline
		(xii) $\text{splitting N\'eel, no large $t_2$}$ & -0.37398 & 0.1339 & 0.4420 & 0.1267 & 0 & 1.0530$(\eta_{\uparrow})$  & 0.0900 & \\ \hline
		(xiii) $\text{splitting N\'eel, no N\'eel order}$ & -0.37392 & 0.6223 & 0.5823 & 0.0582 & 0.1659 & 1.0097$(\eta_{\uparrow})$ & 0 & \\ \hline
	\end{tabular}
	\caption{Energy and variational mean-field parameters for different cases (12$\times12$ sites by default) with $J_1/J_2=0.7, J_r=0$.}
	\label{Tab:MFrules}
\end{table}

{\it (a)The EP phase.}\\
\indent  The EP phase is also a gapless nematic $Z_2$ spin liquid as it does not  breaking translation symmetry. Besides the mirror reflection symmetry breaking parameters $\eta_{x,y}\neq 1$, our results show that the chemical potential $\lambda_z$ is required, while will suppressed by ring exchange term, does not vanish and remains finite (even if it is very small, it is definitely not zero. Because once $\lambda_z$ is not included in the variational process, the energy will increase). Actually, the chemical potential also breaks the mirror reflection symmetry of the Z2Azz13SS PSG since it reflects its sign under the mirror operation. 
 
\begin{figure}[t]
    \centering
        \includegraphics[width=.7\linewidth]{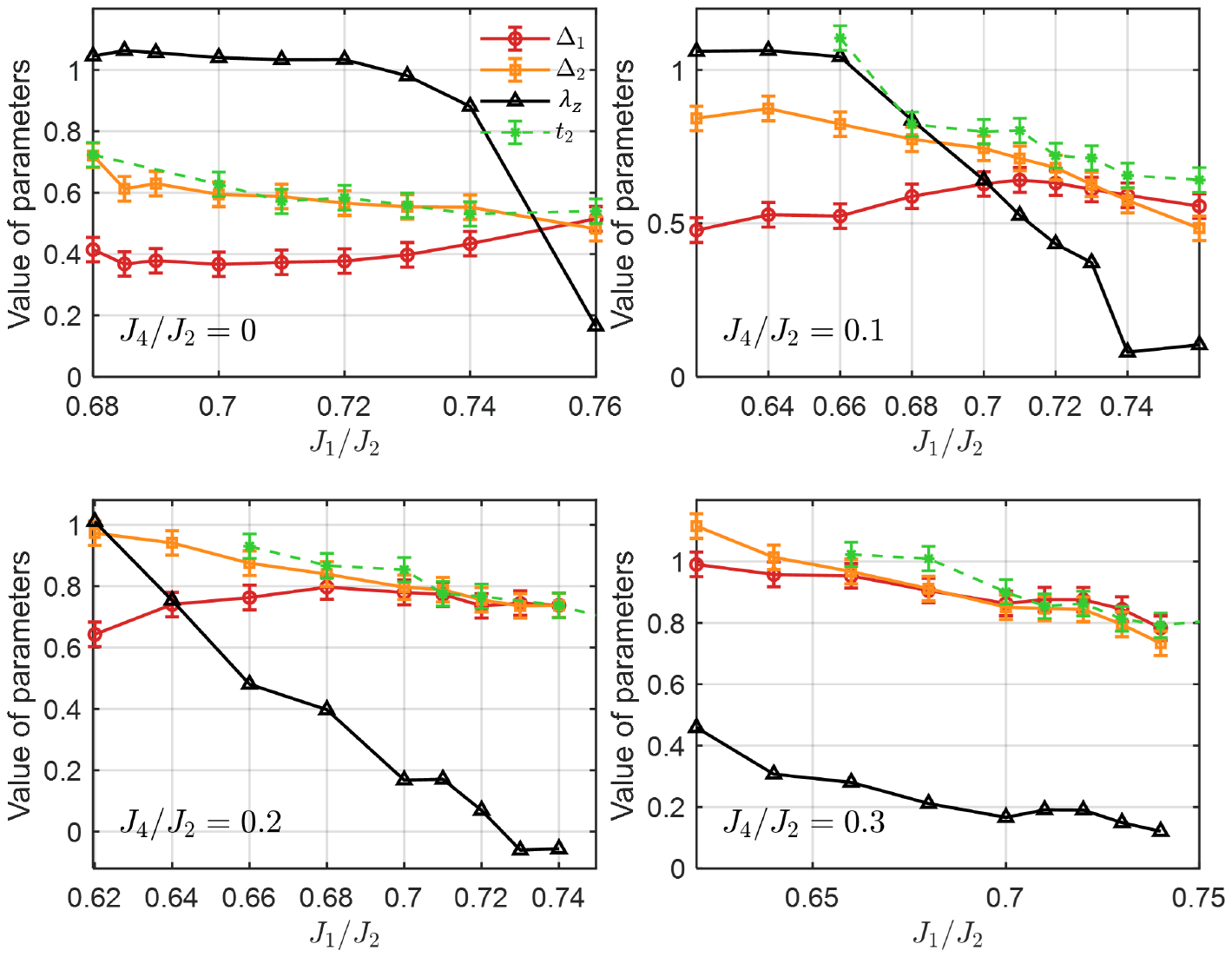}
    \caption{ 
    Evolution of variational parameters in EP and N\'eel phases. The red, orange, and black lines represent the parameters \(\Delta_1\), \(\Delta_2\), and the chemical potential \(\lambda_z\), respectively.  The green line stands  for \(t_2 \)  in the N\'eel phase. The error bars of the chemical potential (not shown) become increasingly large as it approaches zero (estimated to be around $\pm 0.1$), because the energy becomes insensitive to the chemical potential, leading to larger uncertainties in the variational parameters.
      } \label{fig:J4}
\end{figure}

The ansatz in EP phase has $\Delta_2 \neq 0$ and $t_2 \approx 0$ (the detail parameter shown in Tab.\ref{Tab:MFrules} (i) case. In larger system sizes, the energy of this ansatz becomes lower, as in the(vi) case). When $t_2$ slightly deviates from 0,  a different solution with the same energy value can be obtained by amplifying $\Delta_1$ and $\Delta_2$ ((v) case). Furthermore, For $J_r \lesssim 0.2$, $\Delta_2$ is generally greater than $\Delta_1$, as the system approaches the phase transition point, $\Delta_2$ approaches $\Delta_1$, as shown in Fig.\ref{fig:J4}. 

\begin{figure}[h]
	\centering
	\includegraphics[width=0.65\linewidth]{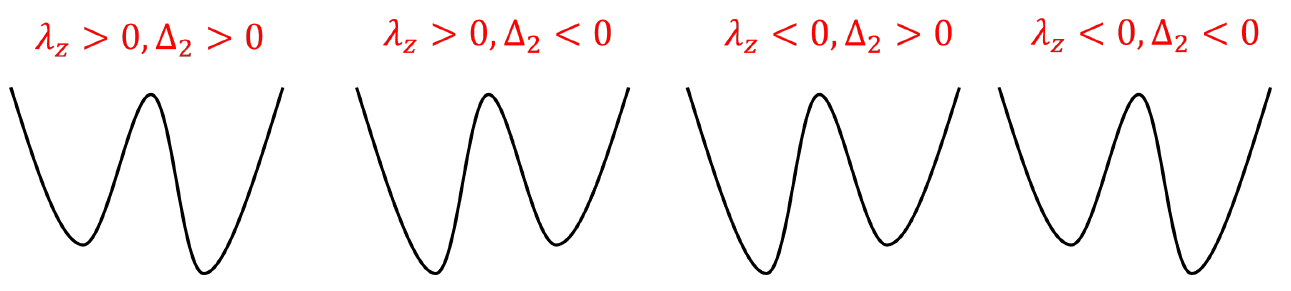}
	\caption{ Schematic sketch of the energy cure in the EP phase, where the minimums stand for the two kinds of EP orders. The horizontal axes represents the variational parameter space. The energies of the two EP orders are the same if $\lambda_z=0$ and split if $\lambda_z\neq0$, therefore $\lambda_z$ effectively provides a `potential energy' for the two EP states.}
	\label{fig:plaquette chose}
\end{figure}

We numerically observed an interesting phenomena((ii)-(iv) case): the pattern of the EP order depends on the sign of $\lambda_z\cdot \Delta_2$, the energy difference depends on the strength of the chemical potential. If one fixes $\lambda_z=0$, we have checked the two EP ordered states have the same energy ( swiching plaquette doesn't change energy for (vii) and (viii) case) or no plaquette order as in the SL phase. But if $\lambda_z\neq 0$, then one of the two EP states are lower in energy, depending on the sign of $\lambda_z$. Hence the chemical potential $\lambda_z$ provides a `potential energy' of the EP order, as illustrated in Fig.\ref{fig:plaquette chose}.

Another important point is that when the initial values of $\Delta_2$ or $t_2$ are not sufficiently large, the variational iteration converges to a state without plaquette order, even if the chemical potential is set to a large value, as illustrated in the (x) case. This indicates that only a sufficiently large $\Delta_2$ can induce the plaquette order, while the chemical potential alone is insufficient. In other words, plaquette order only emerges when the corresponding Higgs field associated with $\Delta_2$ condenses.

{\it (b)Small $\lambda_z$ EP ansatz.}\\
\indent States (a) and (b) are obtained by different choices of the initial value of $\lambda_z$(The blue line in Fig.~\ref{fig:J1J2phase diagram} shows this, corresponding to (vii) and (ix) case in Tab.\ref{Tab:MFrules}). If the chemical potential $\lambda_z$ is small, the variation converges to state (b).  This state exhibits plaquette order (about $0.7<J_1/J_2<0.76$) and N\'eel order (about $J_1/J_2>0.71$) for $J_r=0$. There's a mixed phase of plaquette order and N\'eel order, and because of the lack of chemical potential, the N\'eel order tends to develop early. 

There are two case whose energy are very close, namely (1) (vii) case: $\lambda_z=0$, $t_2\neq0$ $\Delta_2=0$ and (2) (ix) case: $\lambda_z$ is small, $t_2=0$, $\Delta_2\neq0$. Although this state is not the lowest energy state at $J_r=0$, it revealed when the chemical potential is suppressed by the ring exchange, it is more like an unstable intermediate state that provides the transition from $\Delta_2$ rotation to $t_2$. Furthermore, the spin liquid phase which has $\lambda_z=0$ also evolves from this state. Therefore, studying the properties of this state is crucial, as it plays an important role in the emergence of the continuous phase transition.

It should be noted that in the first case the chemical potential is strictly zero, or equivalently, even if a small chemical potential added it does not change the energy. In the second case, the chemical potential is close to zero but never exactly vanishes, so switching the two plaquette orders still have an energy difference. If the chemical potential is not included in the variational parameters ($\lambda_z=0$, showon in (viii) case), the strength of the plaquette order becomes smaller but still remains finite, and the positions of the two plaquette orders can be interchanged without affecting the energy, although the energy is higher, once a small $\mu$ is allowed, come back to (ix) case , the energy decreases and becomes equal to that of the first $t_2$-dominated case. This behavior suggests that $\Delta_2$ also induces a chemical potential.
From the above analysis, we can draw four key points:
(1) the existence of plaquette order is mainly induced by $\Delta_2$(or $t_2$) rather than chemical potential.
(2) the existence of chemical potential on the one hand inhibits the premature N\'eel order, on the other hand, it can reduce energy and further intensify the amplitude of the plaquette order. 
(3) The energy difference between the two plaquette orders depends entirely on the magnitude of the chemical potential $\lambda_z$.
(4) In the EP phase, whenever $\Delta_2$ is finite, a chemical potential necessarily appears. By contrast, for $t_2 \neq 0$, the energy remains unchanged regardless of whether the chemical potential is included or not.


{\it(c)The N\'eel phase.}\\
\indent This phase is characterized by nonzero $t_2$ and background field $M_z$ but vanishing $\Delta_2$ and $\lambda_z$ ((xi) case in Tab.\ref{Tab:MFrules}). When the N\'eel order starts to appear in (b), we find the phenomenon is non trival. VMC calculations indicate that the original ansatz requires a spin-splitting structure of the spinon correction to lower the energy, similar to the altermagnetism observed in magnetic materials, except that here it is the spinons rather than itinerant electrons. The hopping amplitudes of $f_\up$ and $f_\dn$ fermions have different strong-weak modulations (see Fig.~\ref{fig:plaquette}(b)). Moreover, the sign change of $t_2\cdot M_z$ will switch the strong-weak pattern of the hopping amplitudes of the $f_{\up},f_{\dn}$ fermions. 

It is important to emphasize that, analogous to the role of $\Delta_2$ in the EP phase, the N\'eel phase requires both a sufficiently large $t_2$ and a finite $M_z$ to sustain the splitting structure; the absence of either prevents its structure, see (xii) and (xiii) case.Another interesting observation is that $t_2$ equals $\Delta_1$ when the ring-exchange term is not very large (more precisely, in the regime where the transition is first order). However, when approaching the SL region, $t_2$ becomes smaller than $\Delta_1$.

\begin{figure}
	\centering
	\includegraphics[width=.62\linewidth]{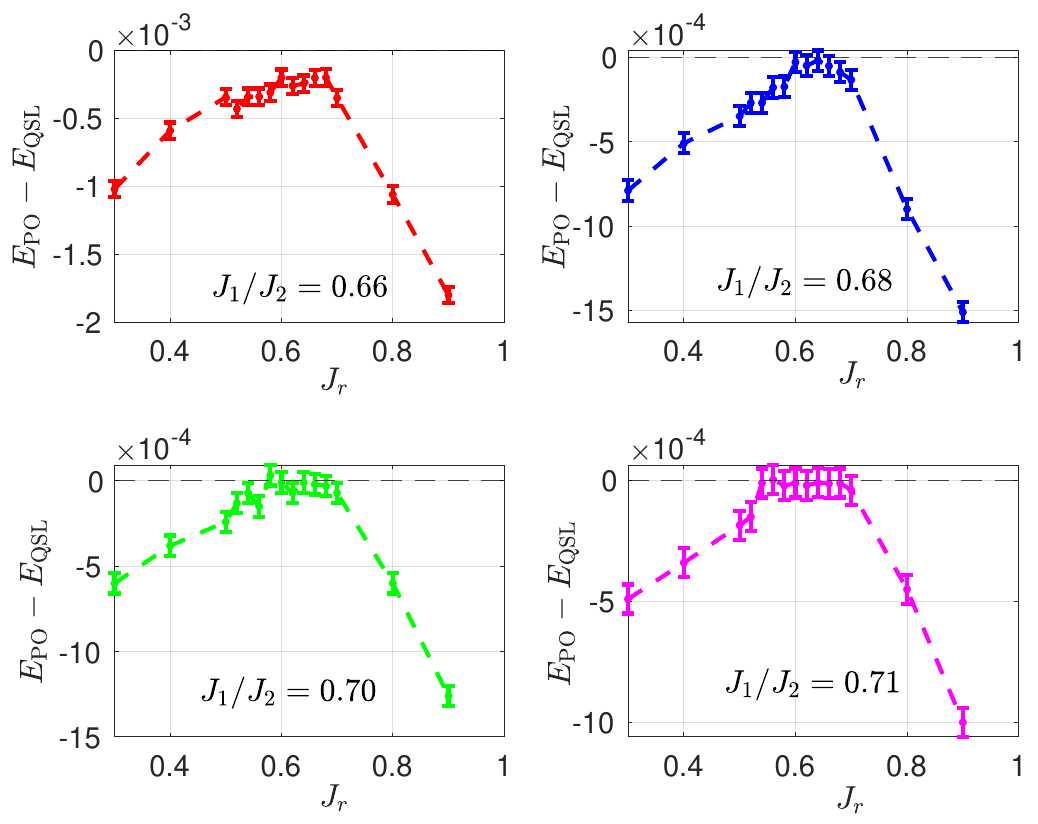}
	\caption{Energy difference $\Delta E= E_{\rm PO}-E_{\rm QSL}$ VS.  \( J_r/J_2\) for fixed $J_1/J_2$, where $E_{\rm PO}$ and $E_{\rm QSL}$ stand for the energies of the state with plaquette orders and of the symmetric state respectively. The region where the energy of the two ansatz converge to the same value is identified as a QSL phase. }
	\label{fig:SL_region}
\end{figure}

{\it (d) The QSL phase.}\\
\indent Within the QSL region, we identified two patterns having the same energy: (1) $t_2 \neq 0$, $\Delta_2 = 0$; and (2) $t_2 = 0$, $\Delta_2 \neq 0$ (see Tab.\ref{Tab:QSLpara}). These two cases have the same energy. This is because, as shown by a sequence of gauge transformations (see App.~\ref{app:Z2Azz13}), $t_2$ and $\Delta_2$ can be rotated into each other , provided that the chemical potential $\lambda_z$ is zero. This equivalent transformation plays a role in facilitating a parameter evolution during the phase transition to the AFM phase. Because the VMC variational parameters exhibit relatively large err (estimated $\pm 0.06$), while the energy variance is much smaller ($\pm 3 \times 10^{-5}$), it is very difficult to locate the phase boundary near the EP and SL phases solely by checking whether the plaquette order $\eta_{x,y}$ equals 1. To overcome this difficulty, we make use of the small energy variance and compare the energy difference between the state with $\eta_{x,y}$ included in the variational parameters  and the QSL state with fixed $\eta_{x,y}=1$. For fixed $J_1/J_2$, we plot the energy difference between the QSL ansatz  ($E_{\rm QSL}$ of a symmetric state with fixed $\eta_{x,y}=1$) and the plaquette ordered states ($E_{\rm PO}$ of a state with $\eta_{x,y}$ determined variationaly) as a function of $J_r/J_2$, as shown in Fig.~\ref{fig:SL_region}. The region in which the two different ansatz yield the same energy is identified as the QSL phase. The N\'eel order $M_z$ has been included in the variational parameters, whose optimal value is vanishingly small (see Tab.\ref{Tab:QSLpara}), indication that a QSL ground state. Our preliminary calculations indicate that at larger system size, the range of the SL phase becomes smaller.

\begin{table}[h]
	\centering
	\label{tab:parameters}
	\begin{tabular}{|c|c|c|c|c|c|c|c|}
		\hline
		$ ansatz $ & E & $t_2$ & $\Delta_1$ & $\Delta_2$ & $\lambda_z$ & $\eta_x$ & $M_z$  \\ \hline
		QSL($t_2\sim0$ gauge) &-0.48169& 0.0702  & 1.0927 &  0.8719 &  0.0474 &1  &  0.0011  \\ \hline
		QSL($\Delta_2\sim0$ gauge) &-0.48165 &0.9044  & 1.1180 & -0.0565 &  0.0171  & 1&  -0.0013  \\ \hline
	\end{tabular}
	\caption{Variational mean-field parameters for two gauge equivalent QSL ansatz for $J_1/J_2=0.71, J_r/J_2=0.6$.}
	\label{Tab:QSLpara}
\end{table}


%

\section{Phase transitions} \label{app:Phasetransitions}
\indent In Variational Monte Carlo analysis, phase transitions can be identified by examining the evolution of variational parameters. Tab.~\ref{table:plaquette} summarizes the values of the order parameters in the vicinity of the phase boundaries. Approaching the SL region, all order parameters show a decreasing trend. However, the magnetic order does not completely vanish at the boundary, which can be ascribed to the neglect of fluctuations in the VMC and finite sizes effect, which become particularly relevant at criticality. This motivates us to place greater emphasis on the evolution of other variational parameters when analyzing the transition. The evolution of the variational parameters, as shown in Fig.~\ref{fig:J4}, reveals that the phase transition tends to evolve from a first-order transition toward a continuous one.

\begin{table} 
	\centering
	\begin{tabular}{|c|c|c|c|c|}
		\hline
					\diagbox{$J_r$/$J_2$}{$J_1/J_2$} & \textbf{0.7} & \textbf{0.71} & \textbf{0.72} & \textbf{0.73} \\
		\hline
		\textbf{0.9} & \textcolor{orange}{1.31} & \textcolor{orange}{1.32} & \textcolor{orange}{1.29} & \textcolor{orange}{1.25} \\
		\hline
		\textbf{0.7} & \textcolor{orange}{1.11} & \textcolor{orange}{1.06} & \textcolor{green}{1.22} & \textcolor{green}{1.21} \\
		\hline
		\textbf{0.6} & \textcolor{black}{1.00} & \textcolor{black}{1.00} & \textcolor{green}{1.17} & \textcolor{green}{1.18} \\
		\hline
		\textbf{0.5} & \textcolor{red}{1.07} & \textcolor{red}{1.04} & \textcolor{green}{1.22} & \textcolor{green}{1.20} \\
		\hline
		\textbf{0.4} & \textcolor{red}{1.11} & \textcolor{red}{1.10} & \textcolor{green}{1.24} & \textcolor{green}{1.23} \\
		\hline
		\textbf{0.3} & \textcolor{red}{1.11} & \textcolor{red}{1.10} & \textcolor{green}{1.27} & \textcolor{green}{1.26} \\
		\hline
		\textbf{0.2} & \textcolor{red}{1.15} & \textcolor{red}{1.14} & \textcolor{green}{1.29} & \textcolor{green}{1.23} \\
		\hline
		\textbf{0.1} & \textcolor{red}{1.21} & \textcolor{red}{1.20} & \textcolor{green}{1.35} & \textcolor{green}{1.35} \\
		\hline
		\textbf{0.0} & \textcolor{red}{1.29} & \textcolor{red}{1.28} & \textcolor{green}{1.42} & \textcolor{green}{1.44} \\
		\hline

	\end{tabular}
	\caption{ Plaquette Strength. These values correspond to $J_1/J_2$ ranging from 0.7 to 0.73 near the phase transition boundary. Red indicates EP, orange is FP, green is splitting AFM, and the green values indicate the strength of the splitting plaquette which is proportional to the strength of the N\'eel order.}
	\label{table:plaquette}
\end{table}

Firstly, we foucs on the \( J_1$-$J_2 \) model with $J_r=0$. As shown in Fig.~\ref{fig:J1J2phase diagram}, the red and green lines represent the EP and N\'eel states, respectively. From the mean-field parameters, the transition between the two phases is a first-order transition because of the sudden change in the chemical potential $\lambda_z$: in the EP phase, \( \Delta_2, \lambda_z \neq 0 \) and \( t_2 \approx 0 \), whereas in the N\'eel phase, \( t_2 \neq 0 \) and \( \Delta_2, \lambda_z = 0 \).

\begin{figure} [h]
	\centering
	\includegraphics[width=.7\linewidth]{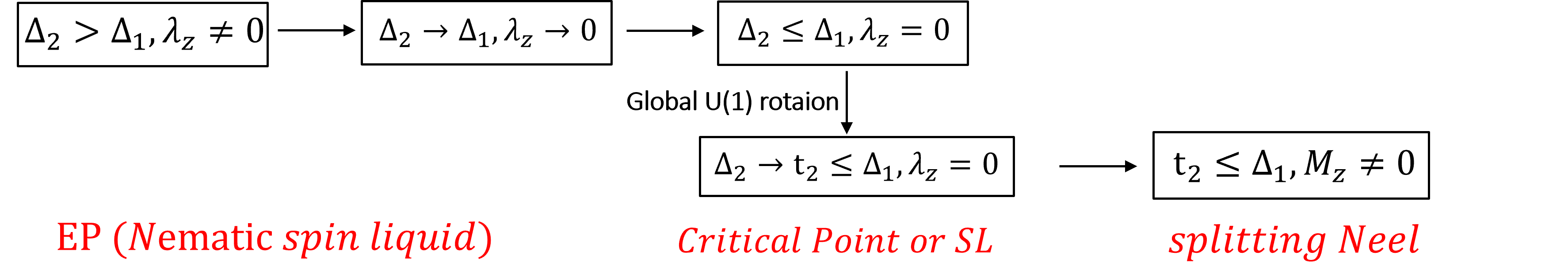}
	\caption{Achieving continuous phase transitions process: EP-AFM or EP-SL-AFM.}
	\label{fig:transation}
\end{figure}

When the ring-exchange term is introduced, the chemical potential in the EP phase are suppressed with increasing $J_r$. The value of \( \Delta_2 \) in the EP phase becomes close to the value of \( t_2 \) in the AFM phase (and both values gradually approach that of \( \Delta_1 \)). Simultaneously, the chemical potential $\lambda_z$ drops to nearly zero. These two events happen almost at the same time (see in Fig.\ref{fig:J4}).  In the gauge (\ref{tilde_uij}), when ignoring the small $\lambda_z$, it can be seen that, through an appropriate gauge transformation, $\Delta_2$ can be rotated into $t_2$. Even in the presence of a small $\lambda_z$, as discussed in App.~\ref{app:mean field rules}, the two energetically close configurations of the small-$\lambda_z$ EP state ($t_2\neq 0$ case or $\Delta_2\neq 0$ case) further support the possibility of such a rotation from $\Delta_2$ to $t_2$. These allowing for a continuous evolution to the AFM phase.  As the N\'eel order develops, spinons split further, giving rise to the splitting N\'eel phase. The evolution of the variational parameters with increasing $J_1/J_2$ is summarized in Fig.~\ref{fig:transation}. Here the "$\leq$" sign refers to the following situation: at a first-order transition, on the N\'eel side one finds $t_2 \approx \Delta_1$, while on the EP side $\Delta_2$ tends to approach $\Delta_1$ (and simultaneously approaches $t_2$). However, when $J_r$ is further increased (where the transition may become continuous or an intermediate SL phase may appear), both $\Delta_2$ and $t_2$ decrease and eventually become smaller than $\Delta_1$. An interesting observation is that when the chemical potential is suppressed to zero, at the onset of the continuous transition occuring (roughly around $J_r \sim 0.3$)—we finds the relation $\Delta_2 = \Delta_1$ in the EP phase and $t_2 = \Delta_1$ in the AFM phase. Although this relation appears somewhat peculiar at first sight, it becomes particularly intriguing from the perspective of the Dirac points. As we know, the Dirac points are located at $(\pi,\pi)$ shifted by 
$
\left( \pm \tfrac{1}{2} \tfrac{\Delta_2}{\Delta_1}, \pm \tfrac{1}{2} \tfrac{\Delta_2}{\Delta_1} \right)
\text{or} 
\left( \pm \tfrac{1}{2} \tfrac{t_2}{\Delta_1}, \pm \tfrac{1}{2} \tfrac{t_2}{\Delta_1} \right).
$
Therefore, the positions of the Dirac points in the EP and AFM phases precisely coincide when approaching the continuous transition point.

\section{Size dependence of the fidelity matrix}\label{app:overlap}

In this appendix, we present the size-dependence of the fidelity matrix $F$ of the 4 wave functions with different boundary conditions. 


In the ideal case, the $Z_2$ deconfinement results in 4-fold degenerate ground states on a torus, while the $Z_2$ confinement yields unique ground state on a torus. In the deconfined phase, the fidelity matrix reads 
$$
F=\Bmat1&&&\\&1&&\\&&1&\\&&&1\Emat,
$$
with eigenvalues $(1,1,1,1)$; while in the confined phase, one has
$$
F=\Bmat 1&1&1&1\\1&1&1&1\\1&1&1&1\\1&1&1&1\Emat,
$$
with eigenvalues $(0,0,0,4)$. 

However, in practice, the eigenvalues deviate from the ideal cases and the eigenvalues of $F$ exhibits are dependent on the system size. To infer the situation at large size limit, we perform calculations for systems with size $8\times8, 12\times12$ and $16\times 16$ and investigate how the eigenvalues evolve with size. 
 If the minimal eigenvalue is much less than 1 and decreases with size, while the maximal eigenvalue is fairly larger than 1 and increases with size, we consider it to be $Z_2$ confined. On the other hand, if the minimal eigenvalue is of order 1 and increases with size, while the maximal eigenvalue is also of order 1 and decreases with size, we consider it to be $Z_2$ deconfined.

The data for the max eigenvalues of $F$ in different phases are listed in Tab.\ref{table:J4}. To distinguish the $Z_2$ confinement or deconfinement, one needs to compare the data in different size and analyze the tendency in large size limit.  

\begin{table}[h]
	\centering
		\begin{tabular}{|c|c|c|c|c|}
			\hline
			\diagbox{$J_r$/$J_2$}{$J_1/J_2$} & \textbf{0.7} & \textbf{0.71} & \textbf{0.72} & \textbf{0.73} \\
			\hline			
			\textbf{0.9} & \textcolor{orange}{2.44} & \textcolor{orange}{2.42} & \textcolor{orange}{2.49} & \textcolor{orange}{2.59} \\
			\hline
			\textbf{0.7} & \textcolor{orange}{2.23} & \textcolor{orange}{2.23} & \textcolor{green}{2.29} & \textcolor{green}{2.22} \\
			\hline
			\textbf{0.6} & \textcolor{red}{2.09} & \textcolor{red}{2.17} & \textcolor{green}{2.24} & \textcolor{green}{2.32} \\
			\hline
			\textbf{0.5} & \textcolor{red}{1.92} & \textcolor{red}{1.93} & \textcolor{green}{2.30} & \textcolor{green}{2.40} \\
			\hline
			\textbf{0.4} & \textcolor{red}{1.91} & \textcolor{red}{1.82} & \textcolor{green}{2.30} & \textcolor{green}{2.40} \\
			\hline
			\textbf{0.3} & \textcolor{red}{1.94} & \textcolor{red}{1.96} & \textcolor{green}{2.31} & \textcolor{green}{2.30} \\
			\hline
			\textbf{0.2} & \textcolor{red}{2.03} & \textcolor{red}{2.01} & \textcolor{green}{2.59} & \textcolor{green}{2.53} \\
			\hline
			\textbf{0.1} & \textcolor{red}{1.25} & \textcolor{red}{1.50} & \textcolor{green}{2.73} & \textcolor{green}{2.73} \\
			\hline
			\textbf{0.0} & \textcolor{red}{1.00} & \textcolor{red}{1.00} & \textcolor{green}{2.88} & \textcolor{green}{2.89} \\
			\hline
		\end{tabular}
	\caption{The max eigenvalue of the fidelity matrix for a system with $12\times 12=144$ sites, 4 means trival state, 1 means 4-fold degeneracy. The red, green, orange stands for the EP, FP and N\'eel phase respectively.}  
	\label{table:J4}
\end{table}

The size-dependence of the fidelity matrix for the SL phase is shown in Tab.\ref{tab:SL_overlap}. The max eigenvalue becomes closer to 2 with the increasing size, indicating the four-fold GSD and $Z_2$ deconfinement. 

\begin{table}[htbp]
	\centering
	\renewcommand{\arraystretch}{1.2} 
	\begin{tabular}{|c|cccc|}
		\hline
		8$\times$8 & 0.0562 & 0.2249 & 1.2208 & 2.4980 \\
		\hline
		12$\times$12 & 0.1577 & 0.3837 & 1.2881 & 2.1705 \\
		\hline
		16$\times$16 & 0.3174 & 0.4897 & 1.2413 & 1.9516 \\
		\hline
	\end{tabular}
	\caption{The SL phase: size-dependence of the eigenvalues of the fidelity matrix for $J_1/J_2 = 0.71$, $J_r/J_2 = 0.6$. The data support 4-fold degenerate ground state on a torus in the large size limit.}
	\label{tab:SL_overlap}
\end{table}

The size-dependence of the fidelity matrix for the EP phase is shown in Tab.\ref{tab:EP_overlap}. The max eigenvalue becomes closer to 1 with the increasing size, indicating the four-fold GSD and $Z_2$ deconfinement. 

\begin{table}[htbp]
	\centering
	\renewcommand{\arraystretch}{1.2} 
	\begin{tabular}{|c|cccc|cccc|cccc|}
		\hline
		${J_r\over J_2}$& \multicolumn{4}{c|}{8$\times$8} & \multicolumn{4}{c|}{12$\times$12} & \multicolumn{4}{c|}{16$\times$16} \\
		\hline
		0.3& 0.1653  & 0.2888 &  1.3097&   2.2362 & 0.3206 & 0.4315 & 1.2797 & 1.9683 & 0.4037 & 0.5610 & 1.1787 & 1.8566 \\
		0.2 &0.3148 &  0.3390 &  1.3791  & 1.9670 & 0.3102 & 0.4039 & 1.2178 & 2.0681 & 0.8097 & 0.9494 & 1.0595 & 1.1815 \\
		0.1&0.3148 &  0.3390 &  1.3791  & 1.9670  & 0.6965 & 0.9686 & 1.0268 & 1.3081 & 0.9466 & 0.9758 & 1.0185 & 1.0592 \\
		0  &0.3234 &  0.3410 &  1.2811 &  2.0546  & 0.9320 & 0.9834 & 1.0189 & 1.0658 & 0.9447 & 0.9772 & 1.0079 & 1.0703 \\		
		\hline
	\end{tabular}
	\caption{The EP phase: size-dependence of the eigenvalues of the fidelity matrix for $J_1/J_2 = 0.7$. The data support 4-fold degenerate ground state on a torus in the large size limit.}
	\label{tab:EP_overlap}
\end{table}

The size-dependence of the fidelity matrix for the FP phase is shown in Tab.\ref{tab:FP_overlap}. The max eigenvalue becomes closer to 4 with the increasing size, indicating the tendency of $Z_2$ confinement. 

\begin{table}
	\centering
	\renewcommand{\arraystretch}{1.2}
	
		\centering
		\begin{tabular}{|c|cccc|cccc|cccc|}
			\hline
			$J_1\over J_2$ & \multicolumn{4}{c|}{8$\times$8} & \multicolumn{4}{c|}{12$\times$12} & \multicolumn{4}{c|}{16$\times$16} \\
		\hline
0.62 & 0.1037 &  0.2522 &  1.2822&   2.3618 & 0.0697  & 0.2411  & 1.0830  & 2.6062&0.0328  & 0.165 &   0.9927 &  2.8087 \\
0.66 &0.0786 &  0.2211 &  1.2390  & 2.4613& 0.0941 &  0.2947&   1.1016 &  2.5096& 0.0478& 0.2140& 1.0080 &2.7302 \\
0.70 &0.0648 &  0.2107 &  1.2188 &  2.5057 &0.0926&   0.2932  & 1.1137 &  2.5005 & 0.0570 & 0.2437 & 0.9937 & 2.7056 \\
\hline
		\end{tabular}
	\caption{The FP phase, size-dependence of the eigenvalues of the fidelity matrix for $J_r/J_2 = 0.9$. The max eigenvalue of the fidelity matrix increase with size, indicating the tendency of single ground state on a torus and $Z_2$ confinement. }	\label{tab:FP_overlap}
\end{table}

The size-dependence of the fidelity matrix for the N\'eel phase is shown in Tab.\ref{tab:AFM_overlap}. The max eigenvalue becomes closer to 4 with the increasing size, indicating the tendency of $Z_2$ confinement. 
\begin{table}
	\centering
	\renewcommand{\arraystretch}{1.2}
	\begin{tabular}{|c|cccc|cccc|cccc|}
		\hline
		${J_1\over J_2}$ &\multicolumn{4}{c|}{8$\times$8} & \multicolumn{4}{c|}{12$\times$12} & \multicolumn{4}{c|}{16$\times$16} \\
		\hline
		0.72 &  0.0903 &  0.1956  & 1.1315 &  2.5825& 0.0280 & 0.0914 & 0.9987 & 2.8819 & 0.0221 & 0.0447 & 1.0013 & 2.9319 \\
		0.73 &0.0657 &  0.1629  & 1.1039  & 2.6675& 0.0259 &  0.0799 &  0.9995&   2.8947& 0.0248 & 0.0420 & 1.0121 & 2.9212 \\
		0.74 & 0.0740 &  0.1650 &  1.1012 &  2.6598 &0.0216 & 0.0674 & 0.9986 & 2.9120 & 0.0139 & 0.0303 & 0.9994 & 2.9564 \\
		\hline
	\end{tabular}
	\caption{In AFM phase: size-dependence of the eigenvalues of the fidelity matrix for $J_r/J_2 = 0$ and varying $J_1/J_2$. The fidelity matrix is not very sensitive to the system size, but its values are clearly greater than 2, which is more like a confined phase.}
	\label{tab:AFM_overlap}
\end{table}

\end{document}